\documentclass[useAMS,usenatbib,usegraphicx,letterpaper]{mn2e}
\usepackage{ gensymb }

\usepackage[fleqn]{amsmath}

\newcommand{\dd}{{\rm d}}
\newcommand{\Pm}{\mathcal{P}}
\newcommand\bnabla{{\bmath\nabla}}
\newcommand\bB{{\bmath B}}
\newcommand\be{{\bmath e}}

\newcommand\rmd{\mathrm{d}}

\newcommand\rms{\mathrm{s}}

\newcommand\p{\partial}

\usepackage{epsfig}
\usepackage{amsmath}

\title[The large-scale magnetic field in protoplanetary discs]
{Global evolution of the magnetic field in a thin disc and its consequences for protoplanetary systems}

\author[Guilet \& Ogilvie]
{J\'er\^ome Guilet$^{1,2}$ and Gordon I. Ogilvie$^1$\\
$^1$ Department of Applied Mathematics and Theoretical Physics, University of Cambridge \\
 Centre for Mathematical Sciences, Wilberforce Road, Cambridge CB3 0WA, UK \\
$^2$ Max-Planck-Institut fur Astrophysik, Karl-Schwarzschild-Str. 1, D-85748 Garching, Germany
}

\begin{document}

\maketitle

\label{firstpage}

\begin{abstract}
The strength and structure of the large-scale magnetic field in protoplanetary discs are still unknown, although they could have important consequences for the dynamics and evolution of the disc. Using a mean-field approach in which we model the effects of turbulence through enhanced diffusion coefficients, we study the time-evolution of the large-scale poloidal magnetic field in a global model of a thin accretion disc, with particular attention to protoplanetary discs. With the transport coefficients usually assumed, the magnetic field strength does not significantly increase radially inwards, leading to a relatively weak magnetic field in the inner part of the disc. We show that with more realistic transport coefficients that take into account the vertical structure of the disc and the back-reaction of the magnetic field on the flow as obtained by Guilet \& Ogilvie (2012), the magnetic field can significantly increase radially inwards. The magnetic-field profile adjusts to reach an equilibrium value of the plasma $\beta$ parameter (the ratio of midplane thermal pressure to magnetic pressure) in the inner part of the disc. This value of $\beta$ depends strongly on the aspect ratio of the disc and on the turbulent magnetic Prandtl number, and lies in the range $10^4-10^7$ for protoplanetary discs. Such a magnetic field is expected to affect significantly the dynamics of protoplanetary discs by increasing the strength of MHD turbulence and launching an outflow. We discuss the implications of our results for the evolution of protoplanetary discs and for the formation of powerful jets as observed in T-Tauri star systems. 
\end{abstract}


\section{Introduction}

The conditions prevailing in protoplanetary discs and their evolution with time are crucial ingredients in the theory of the formation of planetary systems. Magnetic fields have an important impact on the dynamics of protoplanetary discs: they are thought to cause MHD turbulence through the magnetorotational instability \citep{balbus91}, and can launch an outflow through the magnetocentrifugal mechanism \citep{blandford82,ferreira06}. Both of these processes transport angular momentum and therefore largely determine the mass accretion rate and the time-evolution of the disc until its dispersal. The presence of a magnetic field could furthermore directly influence the structure of planetary systems by changing the rate and/or direction of migration of planets embedded in the disc \citep[e.g.][]{terquem03,baruteau11,guilet13a}

The presence of a {\it large-scale} poloidal magnetic field is of particular importance for this dynamics. A strong large-scale poloidal magnetic field is indeed necessary for the launching of a magnetocentrifugal outflow from the inner parts of the disc that could explain the powerful collimated jets observed in T-Tauri star systems \citep{ferreira06}. A weaker poloidal field could also enable the launching of a wind at larger radii that could significantly contribute to driving accretion \citep{suzuki09,fromang13,bai13a,bai13b,suzuki13}. The strength of the large-scale poloidal field is furthermore a key ingredient determining the intensity of MRI-driven turbulence, which is generally more vigorous in the presence of a significant poloidal field \citep{hawley95,bai13a}. Numerical simulations of MRI turbulence in the presence of ambipolar diffusion have even suggested that a net vertical magnetic field in the outer parts of protoplanetary discs is necessary to explain observed mass accretion rates \citep{simon13b}.

Despite its crucial consequences, the strength of the magnetic field remains very uncertain, both from an observational point of view (since direct measurements of its strength are still lacking) and from a theoretical perspective. The evolution of a large-scale magnetic field in an accretion disc has indeed been a long-standing theoretical problem since \citet{lubow94a} and \citet{heyvaerts96} found that the (outward) diffusion of this field was much more efficient than its (inward) advection in a geometrically thin disc. Their analysis was based on a kinematic mean-field approach in which turbulence is modelled by effective diffusion coefficients: a viscosity $\nu$ and a resistivity $\eta$. The radial diffusion of magnetic flux is driven mostly by the vertical diffusion of radial magnetic field across the vertical scaleheight $H$ of the disc, at a typical speed $\eta/H$. On the other hand, advection is driven by viscous transport of angular momentum at a typical speed $\nu/r$ (where $r$ is the radius) if the magnetic flux is assumed to be transported at the same velocity as mass. The ratio of advection to diffusion velocities is therefore $\frac{\nu}{\eta}\frac{H}{r} = \Pm h$ where $\Pm \equiv \nu/\eta$ is the turbulent magnetic Prandtl number and $h=H/r$ is the aspect ratio of the disc. For a realistic magnetic Prandtl number of order unity as expected from MHD turbulence \citep{pouquet76,lesur09,fromang09,guan09}, this ratio is very small if the disc is geometrically thin. As a consequence of this inefficient advection, the magnetic field strength is almost uniform, leading to a negligibly weak magnetic field in the inner parts of the disc, which is problematic for magnetically driven jet models. 

Since then, several ideas have been put forward that could increase the advection speed or decrease the diffusion rate of the magnetic field \citep{spruit05,bisnovatyi-kogan07,guilet12}. It was realized, in particular, that the vertical averaging of the induction equation should not be density-weighted as is usually done for hydrodynamical variables, but rather conductivity-weighted, where the conductivity is the inverse of the effective resistivity \citep{ogilvie01,guilet12,guilet13b}. The vertical structure of the disc (in particular of the resistivity and radial velocity) could therefore significantly change the transport rates compared to the crude estimates of \citet{lubow94a}. \citet{bisnovatyi-kogan07} and \citet{rothstein08} have for example proposed that a non-turbulent surface layer could reduce the diffusion rate of the magnetic field. \citet{guilet12,guilet13b} have performed a radially local calculation of the transport rates taking into account the vertical structure of the disc and the back-reaction of the mean magnetic field on the flow. They have found that a non-turbulent surface layer is ineffective at reducing the diffusion, unless it extends into MRI-unstable regions of the disc which are in fact expected to be turbulent. They also found that for strong magnetic fields such that the magnetic pressure is comparable to the midplane thermal pressure, the estimates of \citet{lubow94a} were a good approximation. On the other hand, for weaker fields they showed that the vertical structure of the disc leads to a faster advection by up to a factor 10 compared to the advection of mass, and a slower diffusion of the magnetic field by a factor up to 4 compared to the estimate of \citet{lubow94a}.  Indeed, the diffusion rate decreases with decreasing magnetic field strength because the magnetic field lines can bend over a larger height, while the advection velocity increases because of the faster radial velocity in the low-density region away from the midplane. This encouraging result suggests that the magnetic field could be efficiently advected and therefore increase radially inwards, which could therefore potentially solve this long-standing problem. Determining the radial structure and intensity of the magnetic field requires however the study of a global model of an accretion disc, which is the subject of this paper.

The theory of star formation suggests that protoplanetary discs may actually form in a highly magnetized state. In fact, if magnetic flux is conserved during the collapse, the magnetic field is too strong and too efficient at removing angular momentum for a rotationally supported disc to form. In the star-formation community the main interest is therefore in how to expel the magnetic flux in order to enable disc formation \citep[e.g.][]{li14}, in contrast to the accretion-disc community which tries to find a way for advection to be more efficient. While part of the magnetic flux may already diffuse out during the collapse itself (for example due to a turbulent resistivity \citep{santos-lima12,joos13}), it is likely that some significant flux remains in the disc when it forms. We will therefore also study the evolution of a protoplanetary disc from a highly magnetized initial condition, and determine on what timescale any initial strong magnetic flux can be diffused out.

To better understand the global structure and time-evolution of the large-scale magnetic field in protoplanetary discs, we study a global one-dimensional model of a thin accretion disc. We use a mean-field approach with the effects of turbulence being modelled by diffusion coefficients and we neglect any additional dynamo effect caused by MRI-driven turbulence. The time-evolution of the magnetic field is determined by transport rates (with contributions from both advection and diffusion) for which we consider different prescriptions. We start by considering simple transport rates following \citet{lubow94a}, which are kinematic and therefore independent of the magnetic field strength. These transport rates do not allow much magnetic field advection if a realistic magnetic Prandtl number is used; however, increasing this parameter to larger values enables us in a simple way to study general properties of the magnetic-field structure in a thin accretion disc when advection is efficient. This part of our study is similar to that of \citet{takeuchi13} and \citet{okuzumi13}, who considered the limit of very efficient advection, except that we also study the regime in which advection and diffusion velocities are comparable. We then use the more realistic transport rates computed by \citet{guilet12}, which take into account the vertical structure of the disc and the back-reaction of the mean magnetic field on the flow. In contrast to the transport rates used previously, these depend on the strength of the magnetic field and we will show that they allow a significant advection weaker magnetic fields for a realistic turbulent magnetic Prandtl number.  

The plan of this paper is as follows. The physical and numerical setup used in this paper is described in Section~\ref{sec:setup}. In Section~\ref{sec:lubow}, we study the time-evolution of the magnetic field when the simple transport coefficients of \cite{lubow94a} are used. We then use the more \textbf{realistic} transport coefficients of \cite{guilet12} in Section~\ref{sec:vertical_structure}. Finally, in Section~\ref{sec:conclusion}, we summarize our results and discuss their consequences for protoplanetary disc evolution and the launching of an outflow.


\section{Physical and numerical setup}
	\label{sec:setup}
\subsection{Formalism}

We study the time-evolution of the large-scale poloidal magnetic field surrounding a thin accretion disc (with a small aspect ratio $H/r = h \ll1$). We use a mean-field approach to describe the time-evolution of the magnetic field, and model the effects of turbulence inside the disc by isotropic effective diffusion coefficients: a viscosity $\nu$ and a resistivity $\eta$. Our neglect of any anisotropy of the diffusion coefficients, as well as of any dynamo $\alpha$ effect that may be generated by MRI-driven turbulence, prevents dynamo action and implies that the evolution of the magnetic field is governed by an advection-diffusion equation. We consider the mean magnetic field averaged spatially in the azimuthal direction and over a time of the order of the dynamical timescale (i.e.\ the Keplerian orbital period). It is therefore axisymmetric and steady on a short dynamical timescale, and we consider its evolution over a longer viscous or resistive timescale.

\subsubsection{Magnetic-field structure}
	\label{sec:B_field_structure}
We assume that no outflow is launched from the disc; therefore outside the disc the density vanishes and the magnetic-field configuration is poloidal and force-free, which implies that it is also current-free. We further neglect the currents present in the star, so that the magnetic field originates only from currents located inside the disc and from currents located at infinity that create a uniform background magnetic field. We describe the poloidal component of the magnetic field with a magnetic flux function $\psi$ as defined by \citet{ogilvie97}. Using a cylindrical coordinate system $(r,\phi,z)$, this magnetic flux function is related to the poloidal magnetic field by
\begin{equation}
\bB = \bnabla \psi \times \be_\phi
\end{equation}
or 
\begin{equation}
B_r = -\frac{1}{r}\frac{\p\psi}{\p z},\qquad
B_z = \frac{1}{r}\frac{\p \psi}{\p r},
	\label{eq:Bz_flux}
\end{equation}
where $\be_\phi$ is the unit vector in the azimuthal direction, while $B_r$ and $B_z$ are the radial and vertical components of the magnetic field. The magnetic flux threading a disc of radius $r$ is proportional to this flux function and equals $2\pi\psi$. This definition of the magnetic flux function $\psi$ is the same as in \citet{ogilvie97} or \citet{heyvaerts96} (where it is called $A$) but differs by a factor of $r$ from that used in \citet{lubow94a}. 

The azimuthal component of the magnetic field is assumed to vanish outside of the disc, in agreement with our assumption that no outflow is launched and therefore that the magnetic torque at the disc surface vanishes. We therefore follow the evolution of the poloidal component of the magnetic field only, which is fully determined by the magnetic flux function\footnote{Note that an azimuthal magnetic field can be present inside the disc and can have dynamical effects that indirectly affect the transport of magnetic flux. This has been taken into account in the analysis of \citet{guilet12}, whose results are used in Section~\ref{sec:vertical_structure}. }. With our assumption of axisymmetry, the poloidal magnetic field comes only from azimuthal currents. The azimuthal current density $j$ is given by
\begin{equation}
\mu_0 j = -\frac{\p}{\p z}\left(\frac{1}{r}\frac{\p\psi}{\p z}\right) - \frac{\p}{\p r}\left(\frac{1}{r}\frac{\p \psi}{\p r} \right).
	\label{eq:def_j}
\end{equation}
Outside of the thin disc, the azimuthal current density vanishes, and the flux function then satisfies the equation
\begin{equation}
\frac{\p}{\p z}\left(\frac{1}{r}\frac{\p\psi}{\p z}\right) + \frac{\p}{\p r}\left(\frac{1}{r}\frac{\p \psi}{\p r} \right) = 0.
	\label{eq:psi_exterior}
\end{equation}

The flux function is separated into a component $\psi_\rmd$ due to currents in the disc and a component $\psi_0$ due to currents at infinity. This second term corresponds to a uniform vertical magnetic field $B_0$ across the whole domain, and therefore 
\begin{equation}
\psi_0 = \frac{r^2}{2}B_0.
\end{equation}  

The flux distribution created by the currents inside the disc can be written as \citep{ogilvie97}
\begin{equation}
\psi_\rmd(r,z) = \frac{1}{\pi}\int_{r_1}^{r_2} \frac{rr^\prime}{\sqrt{(r+r^\prime)^2 + z^2}}\left\lbrack \frac{(2-k^2)K(k) - 2E(k)}{k^2}\right\rbrack J_{\phi}^\rms(r^\prime)\,\dd r^\prime,
	\label{eq:psi_jphi}
\end{equation}
where $k>0$ is defined by
\begin{equation}
k^2 = \frac{4rr^\prime}{(r+r^\prime)^2 + z^2},
\end{equation}
$K$ and $E$ are the complete elliptic integrals of the first and second kind respectively:
\begin{equation}
K(k) = \int_0^{\pi/2}(1-k^2\sin^2x)^{-1/2}\,\dd x,
\end{equation}
\begin{equation}
E(k) = \int_0^{\pi/2}(1-k^2\sin^2x)^{1/2}\,\dd x,
\end{equation}
and $J_{\phi}^\rms$ is the azimuthal current surface density (i.e.\ azimuthal current integrated over $z$). It is related to the radial magnetic field at the surface of the disc, $B_{r\rms}$ (actually the radial magnetic field extrapolated to $z=0^+$ from the exterior force-free field) by $\mu_0J_{\phi}^\rms = 2B_{r\rms}$.

The flux function is therefore fully determined by the radial distribution of azimuthal current density inside the disc, and by the assumed background magnetic field $B_0$. Equation~(\ref{eq:psi_jphi}) can actually be inverted to obtain the azimuthal current density inside the disc $J_{\phi}^\rms$ as a function of the radial distribution of the magnetic flux function in the disc midplane $\psi_\rmd(r,0)$ \citep{lubow94a,ogilvie97}. In the end, the evolution of the 2D magnetic-field structure therefore reduces to a 1D problem where one computes the time-evolution of the magnetic flux function in the disc midplane.

\citet{heyvaerts96} proposed a simpler integral formula to determine $B_{r\rms}$ (or equivalently $J_\phi^\rms$) from $\psi_\rmd$.  Their equation~(126) avoids the need for elliptic integrals and for the inversion of a matrix.  It has been used by them and by subsequent authors \citep[e.g.][]{reynolds06}.  However, we have found that the procedure of \citet{heyvaerts96} is invalid and should not be used.  Their equation~(123) apparently assumes that the variables $r$ and $x$ are nearly equal, which is not true in general; this means that their equation~(126) gives an inaccurate description of the magnetic interaction between different radii in the disc, which should indeed involve elliptic integrals.  Furthermore, their approach requires the flux function to be known for $0<r<\infty$ and allows the poloidal field to bend at the midplane at any radius; it does not allow for the disc to be of finite extent or to have an inner hole in which currents cannot be supported.

\subsubsection{Time-evolution of the magnetic flux}
The time-evolution of the magnetic flux function in the disc midplane is determined by transport processes in the accretion disc, and may be stated in general as
\begin{equation}
\frac{\p \psi}{\p t} + v_{\psi}\frac{\p \psi}{\p r} =\frac{\p \psi}{\p t} + rv_{\psi}B_z= 0,
	\label{eq:flux_time0}
\end{equation}
where $v_\psi$ is the transport velocity of the magnetic flux, which is driven both by the advection of the magnetic field by the accreting matter, and by the diffusion of the magnetic field by an effective turbulent resistivity.  The calculation of the transport velocity $v_\psi$ for a model of a turbulent thin disc is described in \citet{guilet12} and generally involves numerical computations of the vertical structure of the disc; $v_\psi$ depends on the strength, inclination and radial gradient of the magnetic field as well as on several parameters of the disc.  Before implementing these results we discuss analytically some aspects of that problem in order to gain some physical understanding.

The transport velocity of the magnetic flux can be evaluated at any height in the disc as
\begin{equation}
v_{\psi}= v_r + \eta\left( \frac{\p B_r}{\p z} - \frac{\p B_z}{\p r}\right),
	\label{eq:v_psi_z}
\end{equation}
where $v_r$ is the radial velocity.  However, this expression is of limited value in itself because the right-hand side depends on how the field lines bend as they pass through the disc, and also on the vertical profile of the radial velocity, which can be strongly influenced by the magnetic field.  Using a conductivity-weighted average of this equation up to a height $z$ allows one to obtain an expression which does not depend explicitly on the vertical profile of current density but rather on the vertically integrated current density which can be computed from the flux distribution \citep{ogilvie01}:
\begin{equation}
 v_{\psi} = \frac{\bar{\eta}}{z}\left(B_r - z\frac{\p B_z}{\p r} \right) + \bar{v}_r,
 	\label{eq:upsi_average}
\end{equation}
where $\bar{\eta}$ is an average resistivity defined in the following way (it is
actually the inverse of the height-averaged conductivity):
\begin{equation}
\bar{\eta} = \frac{z}{\int_0^z \frac{1}{\eta}\, \dd  z^{\prime}},   \label{eq:eta_average}
\end{equation}
and $\bar{v}_r$ is the average of the radial velocity weighted by the conductivity ($1/\eta$):
\begin{equation}
\bar{v}_r = \frac{1}{\int_0^z \frac{1}{\eta}\, \dd z^{\prime}} \int_0^z \frac{v_r}{\eta}\, \dd z^{\prime}.   \label{eq:vadv_average}
\end{equation}
Note that in a thin accretion disc the vertical magnetic field and therefore the term $\frac{\p B_z}{\p r}$ are independent of height because of the constraint that the magnetic field is divergence free. The term $B_r - z\frac{\p B_z}{\p r}=\int_0^z j \, \dd z^\prime $ corresponds to the vertically integrated current density. This averaging procedure shows that the advection speed of magnetic flux can be different from the advection velocity of mass, since it is conductivity-weighted rather than density-weighted. 

This averaging procedure nevertheless has some drawbacks. First, it depends implicitly on the vertical profile of current density because of the dynamical effect of the magnetic field on the velocity profile. As a result $\bar{v}_r$ cannot be simply identified as the advection velocity of magnetic flux but can also be partly due to a diffusion effect. More importantly, choosing the height up to which the average should be performed is far from obvious. Integrating to `infinity' (i.e.\ up to a height large compared to the scaleheight of the disc) so that $B_r - z\frac{\p B_z}{\p r}$ can be safely identified with the total current surface density (actually to $\mu_0J_{\phi}^\rms/2$) is not a good solution \citep{guilet12}. Indeed, the first term of equation~(\ref{eq:upsi_average}) then vanishes, while the average velocity is dominated by the magnetically dominated region. The result of the average is then the obvious (and useless) result that the radial velocity in the magnetically dominated region equals the advection velocity of the magnetic field. \citet{guilet12} have shown that it is more relevant to perform the average up to the height $z_B$ where the magnetic pressure of the vertical magnetic field equals the thermal pressure. Above $z_B$ the magnetic field becomes force-free such that the current vanishes, while below it the magnetic field is approximately passive such that the radial velocity is that of a hydrodynamical disc. They showed that neglecting the effect of the transition region where the magnetic field is neither passive nor force-free leads to a qualitatively good result, with a difference of order unity with the full solution. In that case the two terms of equation~(\ref{eq:upsi_average}) can be identified as a diffusion due to the current surface density, and an advection velocity. \citet{guilet12} also provided an analytical solution including the effect of the transition region, which is valid in the limit of weak magnetic fields. 

Using an asymptotic expansion of the equations of resitive MHD in the limit of a thin disc and under the assumption of a small inclination of the magnetic field lines with respect to the vertical direction, \citet{guilet12} have shown that the transport velocity of magnetic flux can be decomposed into several terms proportional to gradients in the disc and non-vanishing horizontal magnetic field at the disc surface. Following this decomposition, the transport velocity may be expressed as\footnote{We assume here that the azimuthal magnetic field vanishes at the surface of the disc as explained in Section~\ref{sec:B_field_structure}.}
\begin{equation}
v_\psi = v_{\rm adv} + v_{\rm diff}\frac{B_{r\rms}}{B_z} + v_{DB} \frac{\p \ln B_z}{\p\ln r},
	\label{eq:vphi_decomposition}
\end{equation}
where $v_{\rm adv}$ is the advection velocity of magnetic flux, $v_{\rm diff}$ is the diffusion velocity coefficient resulting from turbulent resistivity acting on the azimuthal current surface density, $B_{r\rms}$ is the radial magnetic field at the surface of the disc (actually the radial magnetic field extrapolated to $z=0^+$ from the exterior force-free field) which is related to the azimuthal current surface density through $B_{r\rms} = \mu_0J_{\phi}^\rms/2$, and finally $v_{DB}$ is a transport velocity coefficient of magnetic flux induced by a radial gradient of vertical magnetic field strength (but no current surface density). Using this decomposition, equation~(\ref{eq:flux_time0}) can then be written as
\begin{equation}
\frac{\p \psi}{\p t} + r\left(v_{\rm adv}B_z + v_{\rm diff}B_{r\rms} + v_{DB} \frac{\p B_z}{\p\ln r}\right) = 0.
	\label{eq:flux_time}
\end{equation}

As already discussed, the transport velocities of magnetic flux $v_{\rm adv}$, $v_{\rm diff}$ and $v_{DB}$ depend on the vertical structure of the disc in a non-trivial way. In this study, we will use two different prescriptions for these transport coefficients. In Section~\ref{sec:lubow}, we use the transport velocities derived by \citet{lubow94a}, and widely used later on \citep[e.g.][]{heyvaerts96,reynolds06}. These transport velocities are kinematic (since they do not take into account the back-reaction of the magnetic field on the flow) and are based on a crude vertical averaging. As noted in the introduction, if a turbulent magnetic Prandtl number of order unity is assumed,  these transport rates do not lead to significant magnetic field advection. Although we do not believe them to be realistic, by artificially increasing the turbulent magnetic Prandtl number we can obtain a simple reference model where general properties of the magnetic-field evolution in a global disc model can be studied.  In Section~\ref{sec:vertical_structure}, we then use the more realistic transport velocities computed by \citet{guilet12}. These transport velocities take into account the vertical structure of the disc and the back-reaction of the mean magnetic field on the flow. We summarize below these two prescriptions for the transport velocities.

In the crude vertical averaging considered by \citet{lubow94a} the transport rates are independent of the magnetic field strength. The advection velocity is assumed to be the same as that of mass (i.e.\ a density-weighted average velocity), which for a steady-state disc far from the inner boundary is
\begin{equation}
v_{\rm adv} = -\frac{3}{2}\frac{\nu}{r},
	\label{eq:vadv_lubow}
\end{equation}
where $\nu$ is the effective turbulent viscosity. The diffusion velocity of the magnetic field is
\begin{equation} 
v_{\rm diff} = \frac{\eta}{H},
	\label{eq:vdiff_lubow}
\end{equation}
where $\eta$ is the effective turbulent resistivity and $H$ is the vertical scaleheight of the disc. Finally, the last term in equation~(\ref{eq:flux_time}) proportional to the radial gradient of vertical magnetic field is neglected as it is expected to be subdominant in a thin disc, i.e.\ one assumes $v_{DB} = 0$.

\citet{guilet12} computed transport rates of the magnetic flux using an asymptotic expansion of the equations of resistive MHD in the limit of a thin accretion disc and of small inclination of the magnetic field lines with respect to the vertical. This allowed them to solve for the vertical structure of the magnetic field and velocity, assuming a simple locally isothermal disc for the vertical thermal structure. The transport velocities appearing in equation~(\ref{eq:flux_time}) can be written in the following way:
\begin{eqnarray}
v_{\rm adv} &=& \left(u_\mathrm{K} + u_{DH}D_H + u_{D\nu\Sigma}D_{\nu\Sigma}\right) \frac{H}{r}c_\mathrm{s},   \label{eq:vadv_vertical_structure}\\
v_{\rm diff}  &=& u_{br\rms} c_\mathrm{s},    \label{eq:vdiff_vertical_structure}\\
v_{DB}        &=&  u_{DB}  \frac{H}{r}c_\mathrm{s},   \label{eq:vdb_vertical_structure}
\end{eqnarray}
where $u_\mathrm{K}$, $u_{DH}$, $u_{D\nu\Sigma}$, $u_{br\rms}$ and $u_{DB}$ are the dimensionless transport  velocity coefficients computed by \citet{guilet12}, and gradients are expressed using the following definitions:
\begin{eqnarray}
D_H &\equiv&\frac{\p\ln H}{\p\ln r}, \\
D_{\nu\Sigma} &\equiv&  \frac{\p\ln\nu\Sigma}{\p\ln r}, \\
D_B &\equiv & \frac{\p\ln B_z}{\p\ln r}.
\end{eqnarray}
In this paper we use the values computed numerically by \citet{guilet12} assuming a turbulent magnetic Prandtl number of $\Pm=1$. These transport velocities depend on the strength of the magnetic field through the dimensionless parameter
\begin{equation}
\beta_0 \equiv \frac{\mu_0}{B_z^2}\frac{\Sigma c_\mathrm{s}^2}{H},
	\label{eq:beta0_def}
\end{equation}
which corresponds roughly to the midplane value of the plasma $\beta$ parameter (the ratio of the thermal pressure to the magnetic pressure); more precisely, the two are related by $\beta(\zeta=0) = \sqrt{2/\pi}\,\beta_0 $.

\subsection{Numerical method}

In order to evolve the magnetic flux function in time, we solve equation~(\ref{eq:flux_time}) using a second-order finite differencing numerical algorithm. The magnetic flux function is discretized on a grid with $n_r$ grid cells distributed logarithmically between $r_\mathrm{min}$ and $r_\mathrm{max}$. The numerical calculations presented in this paper used $n_r=200$. We checked that the results converged with second-order accuracy and that with the fiducial resolution numerical errors remained smaller than $1\%$. 

The time-integration is done using a second-order Runge--Kutta method. The vertical magnetic field involved in the advection term is evaluated from the flux function and equation~(\ref{eq:Bz_flux}) by using a second-order upstream finite differencing. The term involving the derivative of the vertical magnetic field, which has the form a diffusion term, is evaluated using a centred finite differencing. The evaluation of the radial magnetic field at the surface of the disc, related to the current in the disc, deserves further comments. Similarly to \citet{lubow94a}, this is done by writing the integral of equation~(\ref{eq:psi_jphi}) in a matrix form such that the magnetic flux function coming from the currents in the disc is expressed as
\begin{equation}
(\psi_\rmd)_i = \sum_{j=1}^{n_r} Q_{ij}\times(B_{r\rms})_j,
	\label{eq:psi_jphi_matrix}
\end{equation}
where $(\psi_\rmd)_i$ is the value of $\psi_\rmd = \psi - \psi_0$ at the centre of grid cell $i$, $(B_{r\rms})_j$ the value of $B_{r\rms}$ at the centre of grid cell $j$, and $Q_{ij}$ is a matrix yet to be defined. Evaluating the coefficients of $Q_{ij}$ at the centres of the grid cells leads to singular diagonal elements, which was avoided by \citet{lubow94a} by using a smoothing procedure. For better accuracy, we adopt a different method in which the matrix coefficients are obtained by numerically evaluating the following integral over one cell:
\begin{equation}
Q_{ij} \equiv \frac{2}{\pi}\int_{r_{j-1/2}}^{r_{j+1/2}} \frac{r_ir^\prime}{\sqrt{(r_i+r^\prime)^2}}\left\lbrack \frac{(2-k^2)K(k) - 2E(k)}{k^2}\right\rbrack \,\dd r^\prime,
	\label{eq:matrix}
\end{equation}
where $k^2 = \frac{4r_ir^\prime}{(r_i+r^\prime)^2}$, $r_i$ is the radius of the centre of the grid cell $i$, while $r_{j-1/2}$ and $r_{j+1/2}$ are the radii of the inner and outer edges of the grid cell $j$, respectively. The matrix $Q_{ij}$ is then inverted in order to determine $B_{r\rms}$ from $\psi_\rmd$. Note that the matrix $Q_{ij}$, being a property of the grid, needs to be computed and inverted only once. Therefore a more precise determination of the matrix as done here does not increase significantly the computational cost, while it does give significantly better accuracy.

To ensure the stability of the numerical algorithm we use the following CFL condition on the timestep (with a safety factor of $0.5$):
\begin{equation}
\delta t < \min\left(\frac{\delta r}{v_{\rm adv}}, \frac{2\,\delta r}{\pi v_{\rm diff}}, \frac{\delta r^2}{2rv_{DB}}\right),
	\label{eq:cfl}
\end{equation}
where the three terms correspond to the three transport terms of equation~(\ref{eq:flux_time}), and we followed \citet{lubow94a} for the CFL condition of the second term.

A discontinuous profile of the surface current density leads to a logarithmic singularity of the flux function because of the singularity in the integral of equation~(\ref{eq:psi_jphi}). Physically, this comes from the fact that close to an infinitely thin ring current the magnetic field it creates scales like the inverse of the distance from this ring. In order to avoid such a singularity at the inner and outer edges of the disc, we use two buffer zones located in the radius ranges $[r_{\rm min},r_{\rm in}]$ and $[r_{\rm out},r_{\rm max}]$, where the advection velocity goes smoothly from its value in the disc at $r_{\rm in}$ and $r_{\rm out}$ to zero at $r_{\rm min}$ and $r_{\rm max}$. For this purpose, a smoothing function of the following form is used:
\begin{equation}
{\rm smooth}(x) = \frac{3x^2}{1+2x^3},
	\label{eq:smoothing}
\end{equation}
which varies smoothly between 0 for $x=0$ to 1 for $x=1$, with a vanishing derivative at both ends. The variable $x$ is set to $x=(r-r_{\rm min})/(r_{\rm in}-r_{\rm min})$ in the inner buffer zone, and to $x=(r-r_{\rm max})/(r_{\rm out}-r_{\rm max})$ in the outer one.

Finally the last term of equation~(\ref{eq:flux_time}), which has the form of a diffusion term, required a special treatment. Indeed, $v_{DB}$ as computed by \citet{guilet12} is positive, which corresponds to an antidiffusion and can lead to an instability growing on scales smaller than $H$. However, variations of $B_z$ on such length-scales contradict the assumptions made in the analysis of \citet{guilet12}, and the effects of this term at such small scales are not believed to be physical. We have therefore smoothed the second derivative of $B_z$ on a scale of $H$ using a Gaussian function. This removed the instability at small scales observed without this smoothing procedure.


\section{Transport rates independent of magnetic field strength (crude vertical average)}
	\label{sec:lubow}

In this section we describe the time-evolution of the magnetic field calculated using the simple transport rates of \citet{lubow94a}. We use a simple disc model in which the aspect ratio $h=H/r$ is uniform, and the effective viscosity is either uniform or given by an $\alpha$ prescription, $\nu=\alpha H^2\Omega$, with uniform $\alpha$. The effective resistivity is then determined by the magnetic Prandtl number $\Pm$, which is also assumed to be uniform. The transport velocities of the magnetic flux are then obtained from equations~(\ref{eq:vadv_lubow}) and (\ref{eq:vdiff_lubow}). The surface density does not need to be specified as the transport rates of the magnetic flux do not depend on the magnetic field strength through $\beta$ (contrary to Section~\ref{sec:vertical_structure} below). A steady-state disc would however correspond to $\Sigma \propto r^{-1/2}$ for the $\alpha$ disc model, and $\Sigma$ uniform for the uniform viscosity disc model. The disc extends between $r_{\rm in}=1$ and $r_{\rm out}=1000$, and the numerical domain including the buffer zones extends between $r_{\rm min}=0.75$ and $r_{\rm max}=1250$. The magnetic field strength is normalized by the uniform background magnetic field $B_0$.

\subsection{Final stationary state}
	\label{sec:lubow_stationary}
We find that the magnetic-field configuration tends towards a stationary state, which is independent of the radial profile of effective viscosity assumed. The time-evolution towards this stationary state does however depend on the radial profile of the effective viscosity as will be described in Section~\ref{sec:lubow_time_evolution}. First, we describe the properties of the stationary state obtained. As noted by \citet{lubow94a} it depends only on the product of parameters $\Pm h$, which determines the inclination of the magnetic field line at the surface of the disc in a steady state. Indeed, in a steady state the advection and diffusion  of the magnetic flux compensate each other such that
\begin{equation}
\frac{B_{r\mathrm{s}}}{B_z} = -\frac{v_{\rm adv}}{v_{\rm diff}}= \frac{3}{2}\Pm h.
	\label{eq:brs_lubow}
\end{equation}

\begin{figure*}
   \centering
   \includegraphics[width=2\columnwidth]{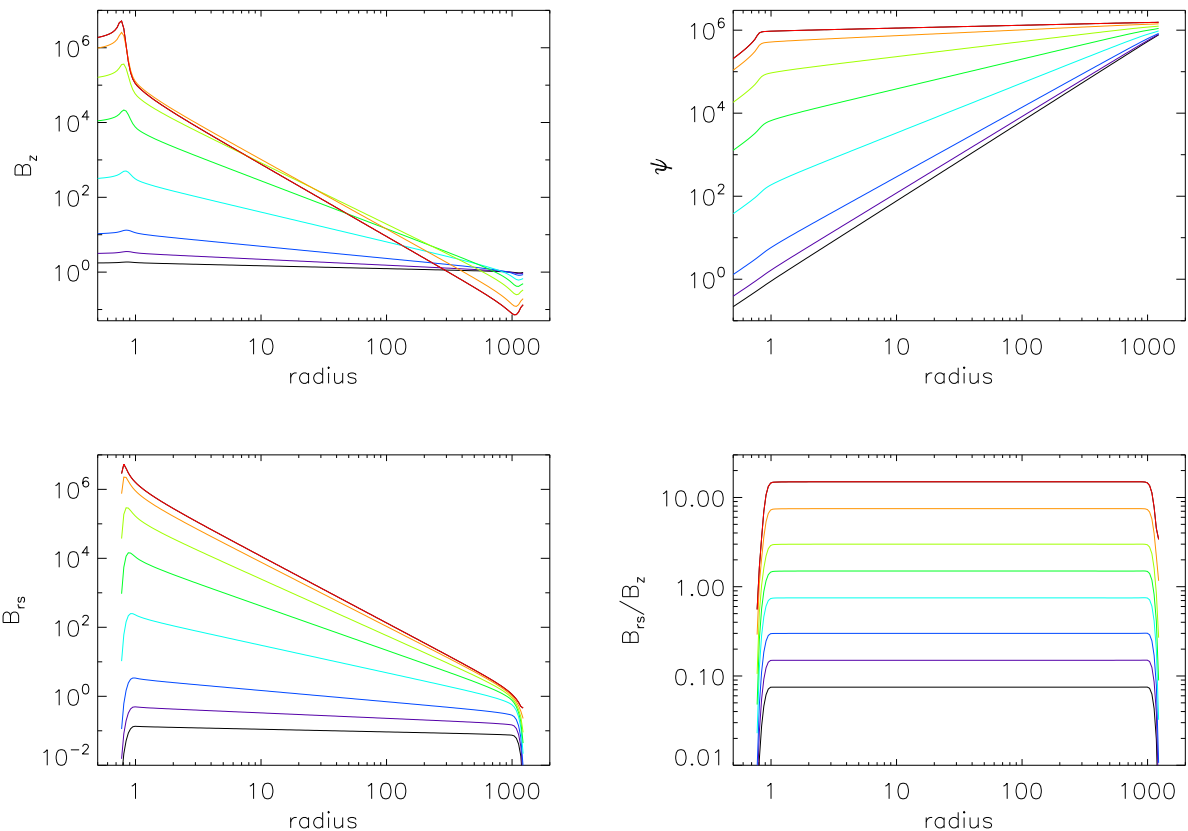}
   \caption{Radial profiles in the stationary state with different values of the parameter $\Pm h$ ranging from $0.05$ to $10$ (shown with colours going from black to red): $0.05$, $0.1$, $0.2$, $0.5$, $1$, $2$, $5$ and $10$. Upper left panel: radial profile of the vertical magnetic field $B_z$. Upper right panel: radial profile of the magnetic flux function $\psi$. Lower left panel: radial profile of the radial magnetic at the surface of the disc, $B_{r\rms}$. Lower right panel: radial profile of the tangent of the inclination of the field line at the surface of the disc, $B_{r\rms}/B_z$.}
   \label{fig:profiles_Pm}
\end{figure*}

\begin{figure*}
   \centering
   \includegraphics[width=\columnwidth]{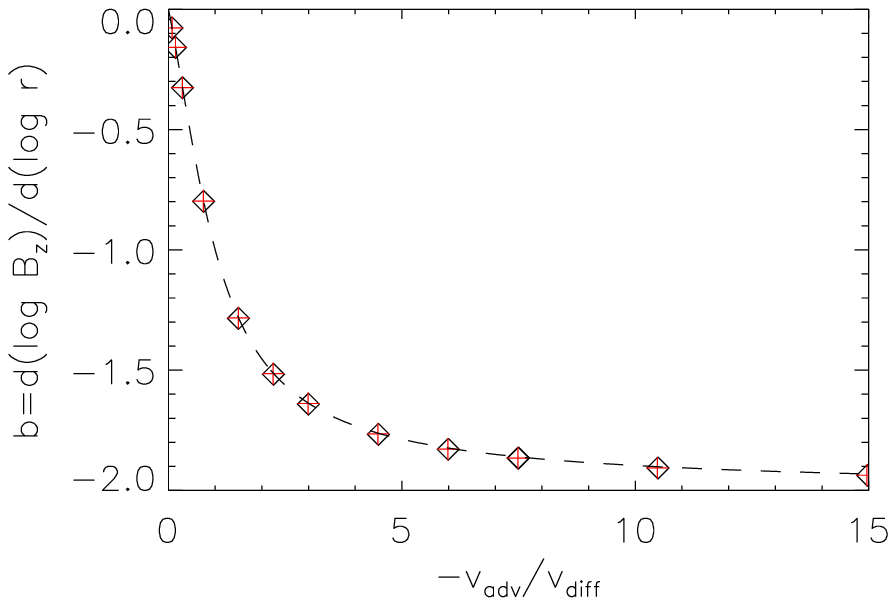}
    \includegraphics[width=\columnwidth]{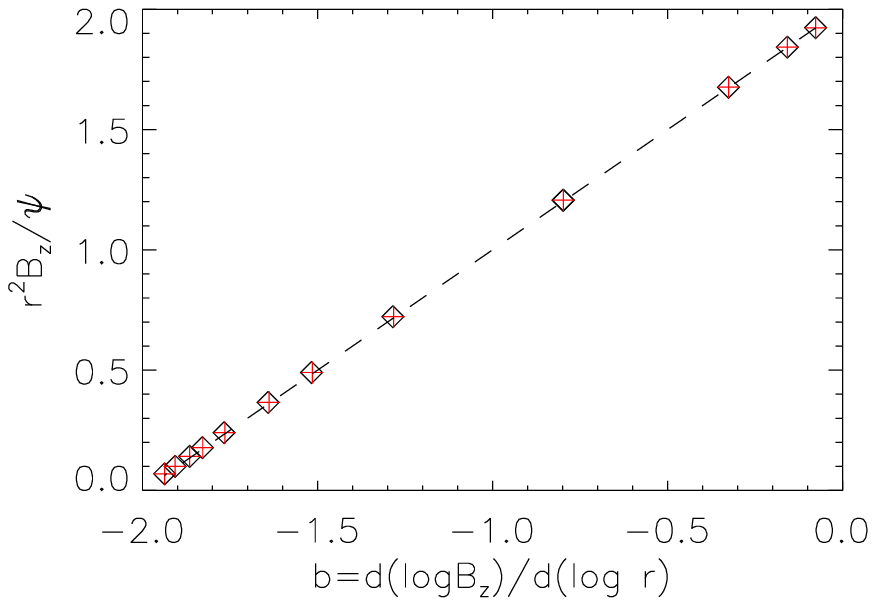}
   \caption{Left panel: Power-law index of the radial profile of vertical magnetic field ($b\equiv  \frac{\p\ln B_z}{\p\ln r}$) as a function of the ratio of advection to diffusion velocities $-\frac{v_{\rm adv}}{v_{\rm diff}}$. The value measured at the end of the simulations in the radius interval $[5,200]$ is shown with diamond symbols for the $\alpha$ disc model and with red $+$ signs for the uniform viscosity disc model. The analytical prediction of equation~\ref{eq:b_power_law} (dashed line) agrees extremely well with the numerical results. Right panel: Relation between the magnetic flux and the vertical magnetic field: $r^2B_z/\psi$ as a function of the power-law exponent $b$. The diamond symbols and red $+$ signs show the result of the simulations (for the $\alpha$ disc model and the uniform viscosity disc model respectively), while the dashed line represents the analytical prediction from the self-similar solution: $r^2B_z/\psi = b+2$ (from equation~(\ref{eq:Bz_flux_relation})).}
   \label{fig:brs_b}
\end{figure*}

Figure~\ref{fig:profiles_Pm} shows the radial profiles of various quantities (vertical magnetic field at the midplane, magnetic flux function at the midplane, radial magnetic field at the disc surface, inclination of the magnetic field lines at the disc surface) in the stationary state and for different values of the parameter $\Pm h$ ranging from $0.05$ to $10$. All these quantities have a power-law dependence on the radius with some deviations close to the boundaries. For small values of the parameter $\Pm h$, the radial inclination of the magnetic field lines is small as expected, and the vertical magnetic field increases inwards only slowly. The magnetic flux is not much affected by the disc and remains close to its initial distribution corresponding to a uniform magnetic field. The inward advection of  magnetic flux is therefore not efficient because the diffusion of magnetic field is more efficient than the advection \citep{lubow94a}. When $\Pm h$ is increased, the radial profile of the vertical magnetic field becomes steeper and steeper, while the magnetic flux distribution departs significantly from the profile corresponding to the background magnetic field. The magnetic flux decreases inwards in a much shallower manner, and the magnetic flux (and magnetic field) in the inner part of the disc is significantly amplified compared to the background magnetic field. The magnetic flux at the outer edge of the disc is increased compared to the flux corresponding to the uniform background magnetic field only by a factor ranging between $1$ (for low $\Pm h$ i.e.\ weak advection) to $2$ (for high $\Pm h$ i.e.\ efficient advection). This interesting upper limit for the increase of the magnetic flux at the outer boundary is consistent with the analytical calculation of \citet{okuzumi13}, which is valid when $\Pm h \gg 1$. 

It is interesting to note that the magnetic field at the outer edge of the disc is always smaller or comparable to the background magnetic field. The current sustained in the disc owing to the magnetic field advection can only increase the magnetic field in the inner part of the disc but not near the outer edge. This can be understood by considering a ring of current located at some radius inside the disc. If the magnetic field is inclined outwards, as expected when diffusion compensates an inward advection, then this current ring acts to increase the vertical magnetic field at smaller radii, and to decrease the magnetic field at larger radii than its own radius. Near the outer edge of the disc, the cumulative effect of all current rings in the disc is dominated by rings located at smaller radii that act to decrease the magnetic field.

Interestingly, we find that, except for the vicinity of the disc edges, the steady-state profiles correspond to the self-similar solution obtained in Appendix~\ref{sec:self_similar} with a power-law dependence of the magnetic field strength with radius. The left panel of Figure~\ref{fig:brs_b} shows the power-law index of the vertical magnetic field $b\equiv  \frac{\p\ln B_z}{\p\ln r}$ as a function of the ratio of advection to diffusion velocities $-\frac{v_{\rm adv}}{v_{\rm diff}}$. The numerical results agree very well with the analytical prediction (obtained by combining equation~(\ref{eq:brs_lubow}) and equation~(\ref{eq:brs_self_similar})):
\begin{equation}
 (b+1)\frac{P_{b+1}(0)}{P_{b+1}^\prime(0)} = -\frac{v_{\rm adv}}{v_{\rm diff}}= \frac{3}{2}\Pm h.
	\label{eq:b_power_law}
\end{equation}
The magnetic field is close to uniform when the advection velocity is small compared to the diffusion velocity, scales like $r^{-1}$ when the two velocities are equal in magnitude, and scales like $r^{-2}$ when the advection velocity is large compared to the diffusion velocity. This is again in agreement with the analytical results of \citet{okuzumi13} who found a maximum magnetic field scaling like $r^{-2}$ in the limit of large advection velocity. 

The self-similar solution predicts a local relation between the magnetic field strength at a given radius and the magnetic flux enclosed inside this radius (equation~(\ref{eq:Bz_flux_relation})). The fact that the magnetic-field profile follows the self-similar solution inside the disc does not guarantee that this relation will also be verified because the magnetic flux depends on the profile of magnetic field down to the origin while the disc is truncated at some inner radius $r_{\rm in}$. The right panel of Figure~\ref{fig:brs_b} shows that this relation is nevertheless verified inside the disc. This means that the magnetic field strength inside the inner disc radius adjusts in such a way as to contain the same magnetic flux as if the power-law distribution of the magnetic field strength were continued down to the centre. This is the reason behind the large values of magnetic field strength at $r<r_{\rm in}$ observed for steep magnetic-field profiles (see the upper left panel of Figure~\ref{fig:profiles_Pm}).

\subsection{Time-evolution towards the stationary state}
	\label{sec:lubow_time_evolution}

\subsubsection{Advection of an initially uniform magnetic field}
	\label{sec:lubow_time_evolution_advection}

\begin{figure*}
   \centering
   \includegraphics[width=2\columnwidth]{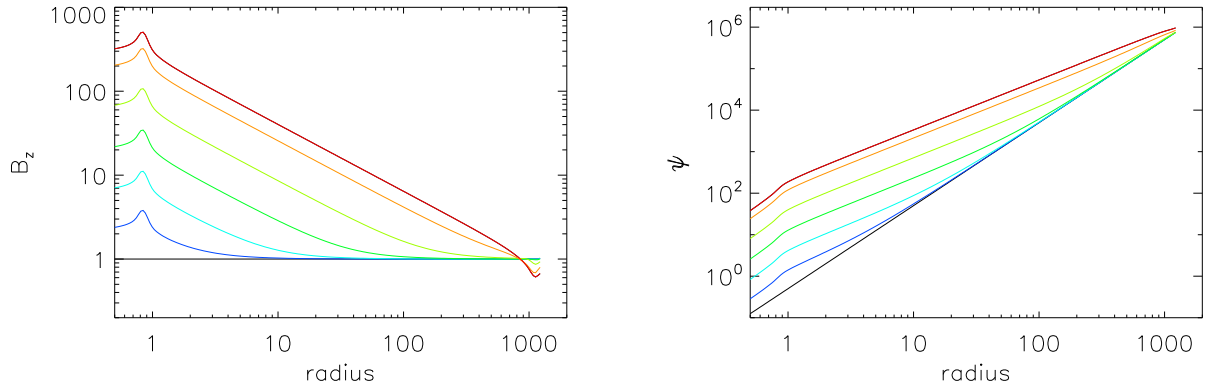}
   \caption{Time-evolution of the radial profiles of vertical magnetic field strength (left panel) and magnetic flux function (right panel). The simulation was initialized with the uniform background magnetic field, and used the parameter $\Pm h = 0.5$. The black line shows the initial condition, while lines with colours going from blue to red show times logarithmically spaced between $t=3.5\times10^{-5}\,t_{\rm visc}(r_{\rm out})$ (approximately the viscous timescale at the inner edge of the disc) and $t=1.5\,t_{\rm visc}(r_{\rm out})$ (by which the profile has reached a stationary state).}
   \label{fig:profiles_time_Pmh1}
\end{figure*}

In this subsection, we study the time-evolution of the magnetic-field profile towards the stationary state, when the initial condition is the uniform background magnetic field (i.e.\ no current is present in the disc). Figure~\ref{fig:profiles_time_Pmh1} shows the time-evolution of the profiles of vertical magnetic field and magnetic flux function for an $\alpha$ disc model with $\Pm h = 0.5$. At intermediate times, the magnetic-field profile shows two separate regions: in the inner disc it has a power-law dependence with the same slope as the final stationary state, while in the outer disc it is still unaffected by the advection and remains uniform. The radius separating these two regions increases with time, and a stationary state is achieved once it has reached the disc's outer radius. The power law present in the inner disc suggests that a quasi-steady state has been reached in this region, in the sense that it is steady compared to the local typical timescale of magnetic flux transport. This quasi-steady state nevertheless evolves on a much longer timescale corresponding to magnetic flux transport at larger radii. 

Following this idea, we propose a simple analytical model to describe this time-evolution. We assume that the inner disc is following the self-similar stationary solution, while the outer disc is still in its initial state. The transition radius between these two regions is estimated by stating that at this radius a significant fraction of the initial magnetic flux has had time to be advected inwards. 
This advection time at a radius $r$ is given by
\begin{equation}
t_{\rm adv}(r) \equiv \frac{r}{v_{\rm adv}} = \frac{2}{3}\frac{r^2}{\nu},
	\label{eq:tvisc_definition}
\end{equation}
which, for an $\alpha$ disc model, can be written 
\begin{equation}
t_{\rm adv}(r) =  \frac{2}{3\alpha h^2\Omega}.
	\label{eq:tvisc_alpha}
\end{equation}
At a given time $t$, the radius such $t_{\rm adv} = t$ is then
\begin{equation}
r_{\rm adv} = r_{\rm out} \left(\frac{3\alpha h^2 \Omega_{\rm out} t}{2} \right)^{2/3} = r_{\rm out} \left(\frac{t}{t_{\rm adv}(r_{\rm out})} \right)^{2/3} ,
	\label{eq:rvisc_alpha}
\end{equation}
where $r_{\rm out}$ is the radius of the outer edge of the disc which we take here as a reference, and $\Omega_{\rm out}$ is the Keplerian angular velocity at that radius. Assuming that the magnetic flux profile is the stationary self-similar solution described in Appendix~\ref{sec:self_similar} for $r<r_{\rm adv}$ and that it connects smoothly to its initial profile ${\textstyle\frac{1}{2}}B_0r^2$ for $r>r_{\rm adv}$, we find the following expression for the magnetic flux at $r<r_{\rm adv}$ :
\begin{equation}
\Psi(r<r_{\rm adv}) = \frac{1}{2}B_0r^2\left( \frac{r}{r_{\rm adv}}\right)^{b}.
	\label{eq:flux_alpha}
\end{equation}
From this equation and equation~(\ref{eq:rvisc_alpha}), one can deduce the value of the magnetic flux at the inner edge of the disc as a function of time :
\begin{equation}
\Psi(r_{\rm in}) = \frac{1}{2}B_0r_{\rm in}^2\left( \frac{r_{\rm in}}{r_{\rm out}}\right)^{b}\left(\frac{t}{t_{\rm adv}(r_{\rm out})} \right)^{-2b/3}.
	\label{eq:flux_in_alpha}
\end{equation}
Using equation~(\ref{eq:Bz_flux_relation}), we obtain the strength of the vertical magnetic field at the inner edge of the disc as a function of time:
\begin{equation}
B_z(r_{\rm in}) =  \frac{b+2}{2}B_0 \left( \frac{r_{\rm in}}{r_{\rm out}} \right)^b\left(\frac{t}{t_{\rm adv}(r_{\rm out})} \right)^{-2b/3}.
	\label{eq:Bzin_alpha}
\end{equation}

\begin{figure*}
   \centering
   \includegraphics[width=2\columnwidth]{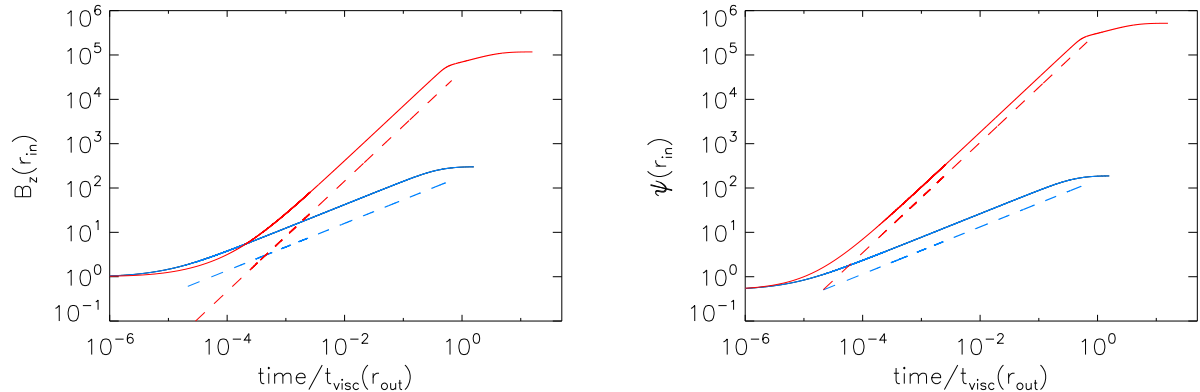}
   \caption{Time-evolution of the vertical magnetic field at the inner edge of the disc (left panel) and of the magnetic flux at the inner edge of the disc (right panel) for the $\alpha$ disc model initialized with the background magnetic field. The time is normalized by the viscous time at the outer edge of the disc, defined as $t_{\rm visc}\equiv r^2/\nu$. The full lines show the result of the simulations with $\Pm h = 0.5$ (blue) and $\Pm h = 5$ (red), while the dashed lines show the corresponding analytical predictions (equations~(\ref{eq:flux_in_alpha}) and (\ref{eq:Bzin_alpha})).}
   \label{fig:flux_Bz_time}
\end{figure*}

\begin{figure}
   \centering
   \includegraphics[width=\columnwidth]{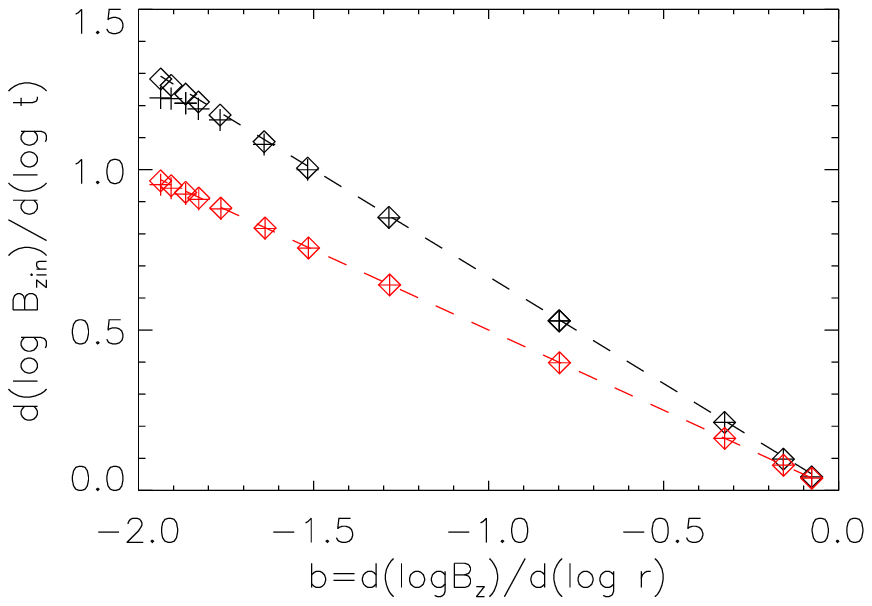}
   \caption{Power-law index of the time-evolution of the magnetic flux at the inner edge of the disc (diamonds) and of the vertical magnetic field at the inner edge of the disc (plus signs) as a function of the power-law index of the radial profile of magnetic field in the stationary state $b$. The analytical predictions are shown with dashed lines. Disc models with a uniform $\alpha$ parameter are shown in black, while those with a uniform viscosity are shown in red.}
   \label{fig:t_power_law}
\end{figure}

These two predictions are compared with the numerical results in Figure~\ref{fig:flux_Bz_time} for two values of the turbulent magnetic Prandtl number $\Pm h=0.5$ and $5$. As predicted, a nice power-law behaviour is observed in the numerical results for both values of the magnetic Prandtl number, indicative of some self-similar behaviour. Furthermore, the power-law index agrees with a good accuracy with the analytical prediction, as shown in Figure~\ref{fig:t_power_law}. The agreement for the normalization of the power law is reasonable (Figure~\ref{fig:flux_Bz_time}), but not very precise, which is not surprising given the simplicity of the model. In particular, the assumption that the magnetic flux function still has its initial value at the radius $r_{\rm adv}$ leads to a slight underestimate since in reality it should already have increased somewhat (similarly to the increase of the magnetic flux function at the outer edge of the disc in the stationary state by a factor in the range $1-2$). The departure from a pure power law near the inner edge of the disc also leads to some discrepancy, in particular in further underestimating the magnetic field.

Following the same reasoning for a disc model with a uniform viscosity, we obtain the following expressions:
\begin{equation}
r_{\rm adv} = \sqrt{\nu t},
	\label{eq:rvisc_uniform}
\end{equation}
\begin{equation}
\Psi(r_{\rm in}) = \frac{1}{2}B_0r_{\rm in}^2\left( \frac{r_{\rm in}}{r_{\rm out}}\right)^{b}\left(\frac{t}{t_{\rm adv}(r_{\rm out})} \right)^{-b/2},
	\label{eq:flux_in_uniform}
\end{equation}
\begin{equation}
B_z(r_{\rm in}) =  \frac{b+2}{2}B_0 \left( \frac{r_{\rm in}}{r_{\rm out}} \right)^b\left(\frac{t}{t_{\rm adv}(r_{\rm out})} \right)^{-b/2}.
	\label{eq:Bzin_uniform}
\end{equation}
A similar agreement with the numerical results is found as for the case of an $\alpha$ disc model. In particular the power-law index, which is different than for the case of an $\alpha$ disc, agrees very well with the prediction as shown in Figure~\ref{fig:t_power_law}. Overall, the results of this section show that an initially uniform magnetic field distribution evolves towards the stationary solution over a timescale set by the advection velocity of the magnetic field.

\subsubsection{Diffusion of an initially steep magnetic-field profile}
	\label{sec:lubow_time_evolution_diffusion}

\begin{figure*}
   \centering
   \includegraphics[width=2\columnwidth]{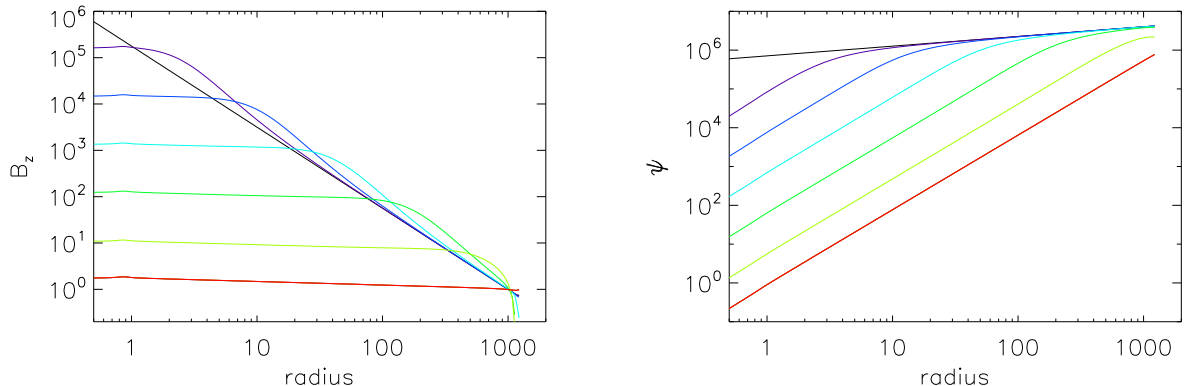}
   \caption{Time-evolution of the radial profiles of vertical magnetic field strength (left panel) and magnetic flux function (right panel). The simulation was initialized with the non-stationary self-similar profile given by equations~(\ref{eq:steep_Bz}) and (\ref{eq:steep_flux}), and used the parameter $\Pm h = 0.05$. The black line shows the initial condition, while lines with colours going from violet to red show times logarithmically spaced between $t=4\times10^{-5}\,t_{\rm visc}(r_{\rm out})$ (approximately twice the resistive timescale at the inner edge of the disc) and $t=1.5\,t_{\rm visc}(r_{\rm out})$ (by which the profile has reached a stationary state).}
   \label{fig:flux_Bz_profiles_time_diffusion}
\end{figure*}

\begin{figure*}
   \centering
   \includegraphics[width=2\columnwidth]{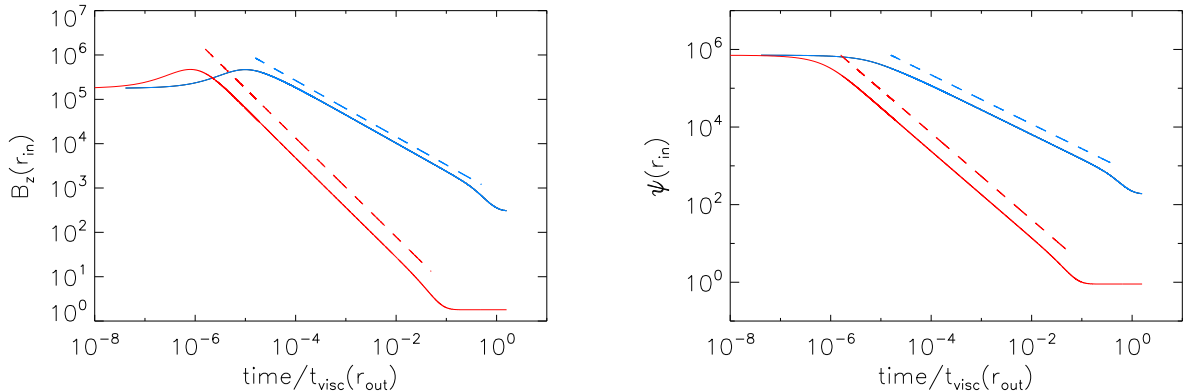}
   \caption{Time-evolution of  the vertical magnetic field at the inner edge of the disc (left panel) and of  the magnetic flux at the inner edge of the disc (right panel) for the $\alpha$ disc model initialized with the steep magnetic-field profile. The time is normalized by the viscous time at the outer edge of the disc. The full lines show the result of the simulations with $\Pm h = 0.5$ (blue) and $\Pm h = 5$ (red). The dashed lines show the analytical prediction (equations~(\ref{eq:flux_in_alpha}) and (\ref{eq:Bzin_alpha})). }
   \label{fig:flux_Bz_time_diffusion}
\end{figure*}

In this subsection, we study the time-evolution of the magnetic-field profile when the initial profile is steeper than the final stationary profile. In that case, the magnetic field is expected to diffuse away, and the timescale over which this happens is relevant to protoplanetary discs as they are likely to form in a rather highly magnetized state. The magnetic-field profile is initialized using the self-similar solution of Appendix~\ref{sec:self_similar} as
\begin{eqnarray}
B_z(r) &=& B_0 \left( \frac{r}{r_{\rm out}} \right)^{bi},    \label{eq:steep_Bz} \\ 
\psi(r) &=& \frac{r^2B_z(r)}{bi+2},    \label{eq:steep_flux}
\end{eqnarray}
where the power-law index of this initial condition is chosen as $bi=-1.75$. 

The time-evolution of the profiles of magnetic field and magnetic flux function is shown in Figure~\ref{fig:flux_Bz_profiles_time_diffusion} in the case $\Pm h = 0.05$ (i.e.\ a realistic magnetic Prandtl number and aspect ratio for protoplanetary discs). At intermediate times, the outward diffusion has decreased the magnetic field in the inner disc, leading to an almost uniform magnetic field in this region, as expected for a (quasi-)stationary solution with this value of the parameter $\Pm h$. In the outer disc the magnetic-field profile is still in its initial state, while in an intermediate range of disc radii the magnetic field is temporarily increased because the magnetic flux initially threading the inner disc has been diffused into this region. As in the last subsection, at intermediate times the profiles show two distinct regions: in the inner disc the profiles have a shape close to the stationary solution (i.e.\ almost uniform magnetic field, magnetic flux scaling like $r^2$), while in the outer disc the profiles remain close to their initial state because diffusion has not yet had time to operate. 

We therefore build an analytical model in a similar vein to the last subsection, where the transition radius is estimated by stating that a significant fraction of the initial magnetic flux has had time to diffuse out. The diffusion time can be estimated as
\begin{equation}
t_{\rm diff}(r) \equiv \frac{r}{v_{\rm diff}} = \frac{r H}{\eta} = \Pm h \frac{r^2}{\nu},
	\label{eq:tdiff_definition}
\end{equation}
which is shorter than the viscous timescale by a factor $\Pm h$. For an $\alpha$ disc model, this can be written
\begin{equation}
t_{\rm diff}(r) =  \frac{\Pm}{\alpha h\Omega}.
	\label{eq:tdiff_alpha}
\end{equation}
At a given time t, the radius such $t_{\rm diff} = t$ is then
\begin{equation}
r_{\rm diff} = r_{\rm out} \left(\frac{\alpha h \Omega_{\rm out} t}{\Pm} \right)^{2/3} = r_{\rm out} \left(\frac{t}{t_{\rm diff}(r_{\rm out})} \right)^{2/3}.
	\label{eq:rdiff_alpha}
\end{equation}
Assuming that the magnetic flux profile is the stationary self-similar solution described in Section~\ref{sec:lubow_stationary} for $r<r_{\rm adv}$ (with a power law index of the magnetic field profile $bf$ given by equation~(\ref{eq:b_power_law})) and that it connects smoothly to its initial profile given by equation~(\ref{eq:steep_flux}) for $r>r_{\rm adv}$, one can deduce the value of the magnetic flux and magnetic field strength at the inner edge of the disc as a function of time :
\begin{equation}
\Psi(r_{\rm in}) = \frac{B_0r_{\rm in}^2}{bi+2}\left( \frac{r_{\rm in}}{r_{\rm out}}\right)^{bf}\left(\frac{t}{t_{\rm diff}(r_{\rm out})} \right)^{-2(bi-bf)/3},
	\label{eq:flux_in_alpha_diff}
\end{equation}
\begin{equation}
B_z(r_{\rm in}) =  \frac{bf+2}{bi+2}B_0 \left( \frac{r_{\rm in}}{r_{\rm out}} \right)^{bf}\left(\frac{t}{t_{\rm diff}(r_{\rm out})} \right)^{-2(bi-bf)/3}.
	\label{eq:Bzin_alpha_diff}
\end{equation}
As shown in Figure~\ref{fig:flux_Bz_time_diffusion}, this analytical model compares favourably with the numerical results, which indeed show a very clear power-law dependence. The numerical results give slightly lower values than the analytical model, which can be interpreted by the fact that the magnetic flux at the transition radius $r_{\rm diff} $ has had time to decrease somewhat.

Overall the results of this section show that any excess magnetic flux initially present in the disc is diffused out on a diffusive timescale given by equation~(\ref{eq:tdiff_definition}). For a thin accretion disc and a turbulent magnetic Prandtl number of order unity, this timescale can be much shorter than the viscous timescale. This is visible in Figure~\ref{fig:flux_Bz_time_diffusion}: for $\Pm h=0.05$, the excess magnetic flux has been expelled and the stationary state is reached within $10\%$ of the viscous timescale at the outer edge of the disc.


\section{Transport rates taking into account the vertical structure of the disc}
	\label{sec:vertical_structure}

\begin{figure*}
   \centering
   \includegraphics[width=2\columnwidth]{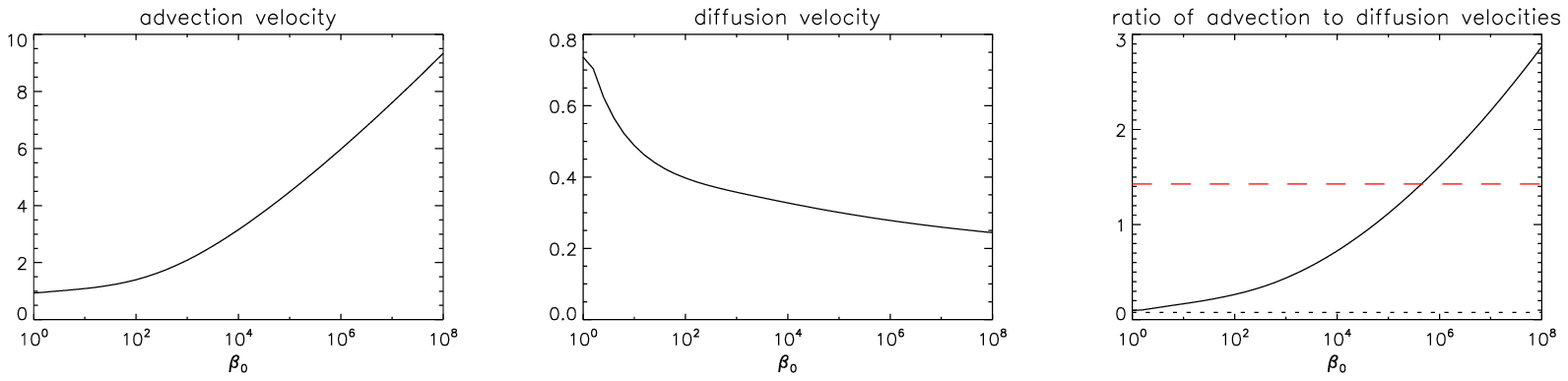}
   \caption{Comparison of the transport velocities of \citet{guilet12} and \citet{lubow94a} as a function of the parameter $\beta_0$ (roughly the ratio of midplane thermal pressure to magnetic pressure). Left panel: advection velocity as computed by \citet{guilet12} divided by the prescription of \citet{lubow94a}. Middle panel: diffusion velocity as computed by \citet{guilet12} divided by the prescription of \citet{lubow94a}. Left panel: ratio of the advection to diffusion velocities for a disc with an aspect ratio of $h=0.05$ and a magnetic Prandtl number of $\Pm=1$. The result of \citet{guilet12} is shown with a full line, while the prescription of \citet{lubow94a} is shown with a dotted line. The red dashed line shows the approximate ratio of transport velocities corresponding to the equilibrium value of $\beta_0$ (see Section~\ref{sec:vertical_structure_stationary}).}
   	\label{fig:transport_velocities}
\end{figure*}

In this section, we study the time-evolution of the magnetic field calculated by using the transport velocities obtained by \citet{guilet12}\footnote{Note that in their calculation, the value of $\alpha$ is computed by assuming marginal stability with respect to the largest-scale MRI mode, which gives larger values of $\alpha$ than what is believed to be realistic. Nevertheless all the transport velocities (advection and diffusion velocity of magnetic field, advection velocity of mass) are expected to be proportional to $\alpha$, such the ratio of these velocities remains relevant. As a consequence the stationary magnetic-field profile (which depends mostly on the ratio of advection to diffusion velocity of the magnetic field) as well as the ratio of the diffusion time to the viscous timescale (studied in Section~\ref{sec:vertical_structure_dissipation}) should not be affected by these unrealistic values of $\alpha$.}. Their analysis takes into account the vertical structure of the disc and the back-reaction of the magnetic field on the flow. As a result the transport velocities differ significantly from those of \citet{lubow94a} used in Section~\ref{sec:lubow}, and they depend on the strength of the magnetic field. Figure~\ref{fig:transport_velocities} compares the advection and diffusion velocities of \citet{guilet12} with those of \citet{lubow94a} for varying magnetic field strength (parameterized by $\beta_0$ defined by equation~(\ref{eq:beta0_def}), which is roughly the ratio of the midplane thermal pressure to the magnetic pressure). The two prescriptions roughly agree for a strongly magnetized disc with $\beta_0 \sim 1$. As the magnetic field strength decreases ($\beta_0$ increases), however, the advection velocity predicted by \citet{guilet12} increases significantly, while the diffusion velocity decreases. As a result the ratio of advection to diffusion velocities (which sets the stationary magnetic-field profile; see Section~\ref{sec:lubow}) can be larger than that predicted by \citet{lubow94a} by up to a factor of $50$. As an example relevant to protoplanetary discs, for a turbulent magnetic Prandtl of $\Pm=1$ and a disc aspect ratio of $h=0.05$, \citet{lubow94a} predict a very inefficient advection with $-v_{\rm adv}/v_{\rm diff}=0.075$ while \citet{guilet12} find that $-v_{\rm adv} > v_{\rm diff}$ if the magnetic field is sufficiently weak with $\beta_0 > 10^5$.

Note that the transport velocities of \citet{guilet12} used here are based on the assumption that the diffusion coefficients are independent of height in the disc. \citet{guilet13b} studied different vertical profiles of the diffusion coefficients, and found that the ratio of advection to diffusion velocities of the magnetic flux was not much affected (both for a fully turbulent disc and a disc containing a dead zone). Since the stationary state of the magnetic field mainly depends on this ratio, we expect that the results presented in Section~\ref{sec:vertical_structure_stationary} would not be much affected by considering a more complicated vertical structure of the diffusion coefficients. The time it takes to reach this stationary state may, however, be somewhat different with the transport coefficients of \citet{guilet13b} as will be discussed in Section~\ref{sec:vertical_structure_dissipation}.

We consider a simple disc model with a uniform aspect ratio, taken in the range $h=0.03-0.1$, orbiting around a solar-mass star. The magnetic Prandtl number is set to $\Pm=1$, which is expected to be representative of MHD turbulence \citep{pouquet76,lesur09,fromang09,guan09}. The surface-density profile is held fixed and corresponds to a steady-state $\alpha$ disc model far from the boundaries: $\Sigma = 50\, (r/10 \, {\rm AU})^{-1/2}\, {\rm g\,cm^{-2}}$. This would correspond to a mass accretion rate of $10^{-7}\,{\rm M_\odot\,yr^{-1}}$ if $h=0.1$ and $\alpha=10^{-2}$. The disc extends between $r_{\rm in} = 0.1\, {\rm AU}$ and $r_{\rm out}=100\,{\rm AU}$, and the numerical domain including the buffer zones extends between $r_{\rm min} = 0.075\,{\rm AU}$ and $r_{\rm max}=125\,{\rm AU}$.

Note that this particular normalization of the surface-density profile is chosen only as a reference model of a protoplanetary disc, but because the only relevant parameter in our setup is the magnetization $\beta_0$ (in addition to the aspect ratio and turbulent magnetic Prandtl number), the results can be rescaled to obtain the magnetic-field profiles corresponding to any other normalization of the surface-density profile by using the relation
\begin{equation}
B_z = \frac{1}{r}\sqrt{\frac{\mu_0\Sigma h G M_\odot}{\beta}}
	\label{eq:Bz_renormalization}
\end{equation}

\subsection{Stationary magnetic-field profile}
	\label{sec:vertical_structure_stationary}

\begin{figure*}
   \centering
   \includegraphics[width=2\columnwidth]{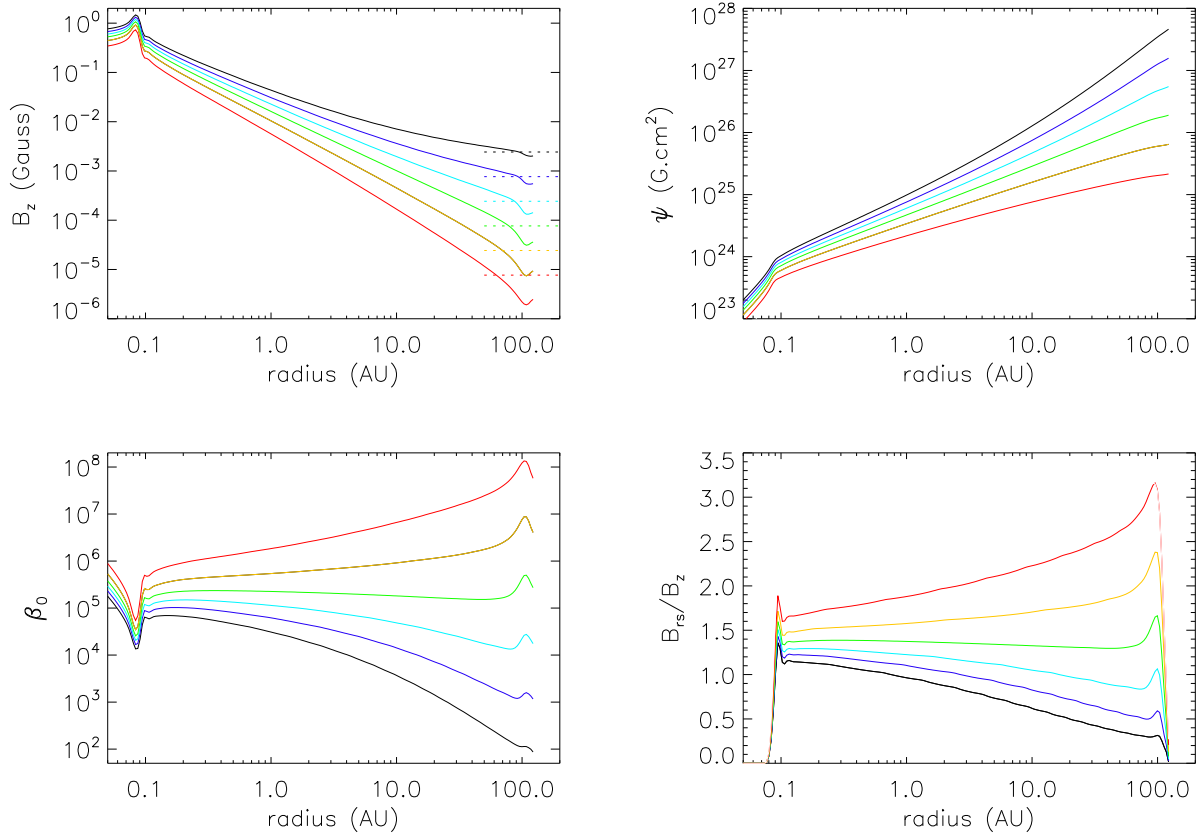}
   \caption{Radial profiles in the stationary state with an aspect ratio of $h=0.05$, for different values of the background magnetic field (shown with colours going from black to red) corresponding to a value of $\beta_0$ at the outer edge of the disc of $10^{2}$, $10^{3}$, $10^4$, $10^5$, $10^6$ and $10^7$. Upper left panel: radial profile of the vertical magnetic field $B_z$. The value of the uniform background magnetic field is shown for comparison with dotted lines. Upper right panel: radial profile of the magnetic flux function $\psi$. Lower left panel: radial profile of the parameter $\beta_0$ (roughy the ratio of midplane thermal pressure to magnetic pressure). Lower right panel: radial profile of the tangent of the inclination of the field line at the surface of the disc, $B_{r\rms}/B_z$.}
   \label{fig:vertical_structure_stationary_profile}
\end{figure*}

The magnetic-field profile evolves towards a stationary state, which is independent of the initial profile. In this subsection we describe this stationary state and leave until the next subsection the description of the time-evolution from the initial state. With our choice of a turbulent magnetic Prandtl number of $\Pm=1$, the stationary magnetic-field profile depends only on the aspect ratio of the disc and on the uniform background magnetic field. In Figure~\ref{fig:vertical_structure_stationary_profile}, we show the stationary profile of several quantities for an aspect ratio of $h=0.05$, and for different values of the background magnetic field. In contrast to Section~\ref{sec:lubow}, the magnetic field does increase significantly inwards for a realistic turbulent magnetic Prandtl number (the curve corresponding to the same aspect ratio and turbulent magnetic Prandtl number in Figure~\ref{fig:profiles_Pm} is the black line, which is almost flat). This was expected because the ratio of advection to diffusion velocities is larger than unity if the magnetic field is weak enough (see Figure~\ref{fig:transport_velocities}). In the case corresponding to the weakest background magnetic field, the magnetic field strength increases by almost five orders of magnitudes between the outer and the inner edge of the disc. For stronger background magnetic fields, the magnetic field strength increases less steeply inwards, as could be expected from the fact that the ratio of advection to diffusion velocities is smaller for stronger magnetic fields. It is indeed clear from Figure~\ref{fig:vertical_structure_stationary_profile} that smaller values of $\beta_0$ correspond to smaller inclinations of the magnetic field at the disc's surface and to less steep magnetic-field profiles, as could be expected from the results of Section~\ref{sec:lubow} and from the dependence of the ratio of advection to diffusion velocities on $\beta_0$.

It is interesting to note that the magnetic field strength at the outer edge of the disc is comparable to or smaller than the background magnetic field. As in Section~\ref{sec:lubow}, the steepest magnetic-field profiles correspond to the smallest ratios of the magnetic field at the disc's outer edge to the background magnetic field ($\sim 1/3$ for the weakest background magnetic field considered here). Also in agreement with Section~\ref{sec:lubow}, the magnetic flux at the disc's outer edge is a factor $1$--$2$ larger than the magnetic flux corresponding to the uniform background magnetic field.

An important difference with the results of Section~\ref{sec:lubow} is that, because of the dependence of the transport velocities on the magnetic field strength,  the magnetic-field profile is not generally speaking a power law. Indeed, for strong background magnetic fields, the magnetic field has a shallow inward increase, such that $\beta_0$ increases inwards. As a consequence, advection becomes more efficient and the magnetic-field profile is steeper in the inner parts of the disc than in the outer parts. Conversely, for weak background magnetic fields, the magnetic-field profile is steep and $\beta_0$ decreases inwards, such that the magnetic-field profile in the inner parts of the disc is shallower than in the outer parts. In between these two regimes lies an equilibrium value of $\beta_0$ for which the slope of the magnetic-field profile is such that $\beta_0$ is uniform. In that case, the magnetic-field profile is therefore a self-similar power law similar to the results of Section~\ref{sec:lubow}. For the parameters used in Figure~\ref{fig:vertical_structure_stationary_profile}, this equilibrium value is $\beta_0 \simeq 2\times10^5$ (green curves). Interestingly, whatever the strength of the background magnetic field (setting the value of $\beta_0$ at the outer edge of the disc), $\beta_0$ tends towards this equilibrium value in the inner parts of the disc, though it may not always have enough radial range to reach it closely.

\begin{figure}
   \centering
   \includegraphics[width=\columnwidth]{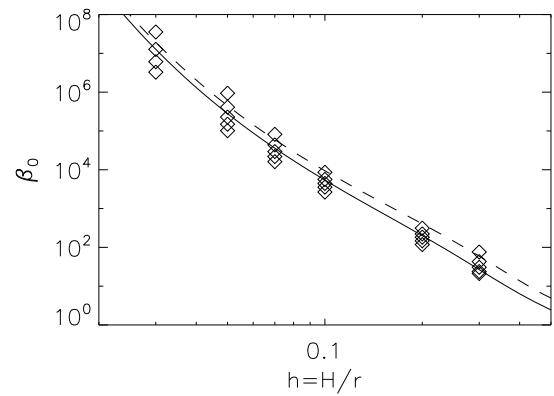}
   \caption{Ratio of thermal pressure to magnetic pressure ($\beta_0$) near the inner edge of the disc as a function of the disc's aspect ratio. The results of the numerical simulations with different initial magnetic field strength are shown with diamond symbols. The analytical prediction of equation~(\ref{eq:beta_eq1}) is shown with the full black line, while the dashed line shows the analytical estimate neglecting the subdominant term $v_{DB}$ (equation~(\ref{eq:beta_eq2})).}
   \label{fig:betaeq_h}
\end{figure}

The self-similar profile corresponding to a uniform $\beta_0$ can be described analytically in the following way. The constraint that $\beta_0$ be uniform determines the power-law index of the magnetic field to be
\begin{equation}
b = -1 + \frac{1}{2}\frac{\p \ln \Sigma}{\p \ln r},
	\label{eq:b_uniform_beta}
\end{equation}
which gives $b=-1.25$ for the surface-density profile considered here. The self-similar exterior magnetic field solution described in Appendix~\ref{sec:self_similar} determines the inclination of the magnetic field at the disc's surface via equation~(\ref{eq:brs_self_similar}), which gives $B_{r\rms}/B_z = 1.428$ for our surface-density profile. The equilibrium value of $\beta_0$ can then be obtained by using equation~(\ref{eq:flux_time}) in a stationary state:
\begin{equation}
v_{\rm adv}B_z + v_{\rm diff}B_{r\rms} + v_{DB} \frac{\p B_z}{\p\ln r} = 0,
	\label{eq:flux_time_stationary}
\end{equation}
which after using equation~(\ref{eq:brs_self_similar}) gives
\begin{equation}
 \frac{B_{r\rms}}{B_z} = (b+1)\frac{P_{b+1}(0)}{P_{b+1}^\prime(0)}= -\frac{v_{\rm adv}+ v_{DB} b}{v_{\rm diff}}.
	\label{eq:beta_eq1}
\end{equation}
The term $v_{DB}$ is subdominant because it is smaller than $v_{\rm diff}$ by a factor of order $H/r$  (see equations~(\ref{eq:vdiff_vertical_structure}) and (\ref{eq:vdb_vertical_structure})), such that one can obtain the simpler approximate expression
\begin{equation}
 (b+1)\frac{P_{b+1}(0)}{P_{b+1}^\prime(0)} = -\frac{v_{\rm adv}}{v_{\rm diff}}.
	\label{eq:beta_eq2}
\end{equation}

These analytical predictions are compared successfully with the results of numerical calculations in Figure~\ref{fig:betaeq_h}, which shows the equilibrium value of $\beta_0$ as a function of the aspect ratio of the disc. In the numerical calculations, the equilibrium $\beta_0$ is estimated by measuring $\beta_0$ near the inner edge of the disc for different values of the background magnetic field (diamond symbols). The equilibrium value of $\beta_0$ depends steeply on the aspect ratio of the disc. For aspect ratios typical of protoplanetary discs ($h\sim0.03-0.1$) and our assumed turbulent magnetic Prandtl number of $\Pm=1$, it lies in the range $10^4-10^7$. 

Note that since the ratio of advection to diffusion velocities is expected to depend mainly on the aspect ratio of the disc $h$ and the turbulent magnetic Prandtl number $\Pm$ through the parameter $\Pm h$ \citep{guilet12}, one would expect the same to be true for the equilibrium value of $\beta_0$. The dependence on the rather uncertain turbulent magnetic Prandtl number is therefore expected to be quite steep as well. For example, assuming $\Pm=2$, one would instead obtain a range of equilibrium $\beta_0$ of $10^2-10^5$ for protoplanetary discs with $h \sim 0.03-0.1$.

\subsection{Time-evolution from a highly magnetized initial condition}
	\label{sec:vertical_structure_dissipation}

\begin{figure*}
   \centering
   \includegraphics[width=2\columnwidth]{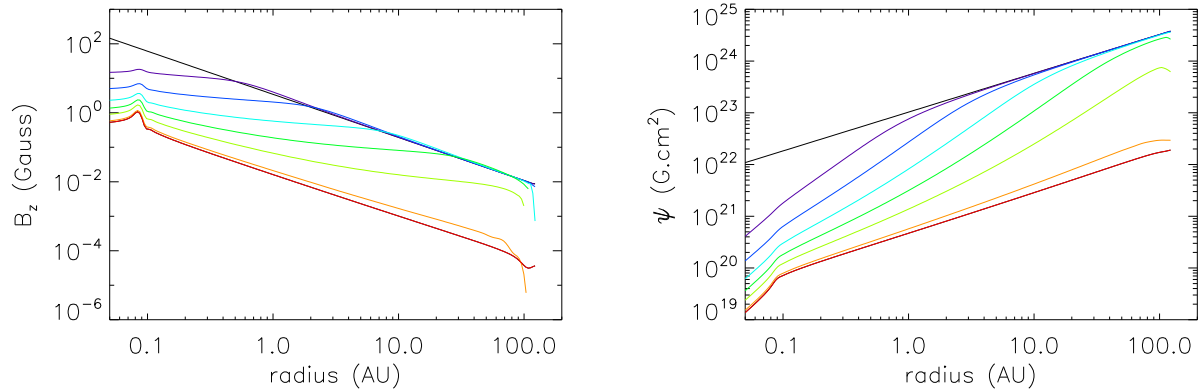}
   \caption{Time-evolution of the radial profiles of the vertical magnetic field strength (left panel) and of the magnetic flux function (right panel). The simulation was initialized such that $\beta_0=5$ uniformly in the disc, and we used an aspect ratio $ h = 0.05$. The black line shows the initial condition, while lines with colours going from violet to red show times logarithmically spaced between $t=3\times10^{-5}\,t_{\rm visc}(r_{\rm out})$ (approximately the viscous timescale at the inner edge of the disc) and $t=4\,t_{\rm visc}(r_{\rm out})$ (by which the profile has reached a stationary state).}
   \label{fig:profiles_time_dissip_vertical_structure}
\end{figure*}

\begin{figure*}
   \centering
   \includegraphics[width=2\columnwidth]{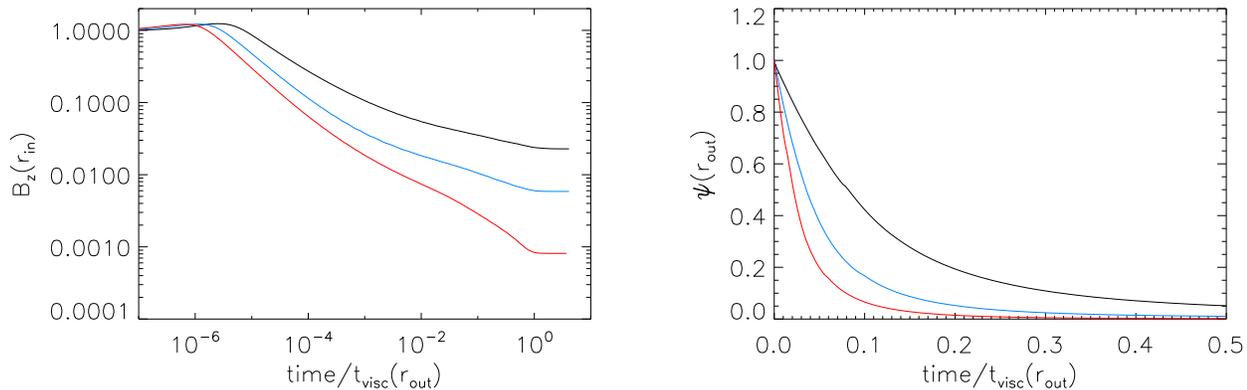}
   \caption{Time-evolution  of the vertical magnetic field at the inner edge of the disc (left panel) and of the magnetic flux at the outer edge of the disc (right panel) in simulations initialized such that $\beta_0=5$ uniformly in the disc. Both quantities are normalized by their initial values, while time is normalized by the viscous time at the outer edge of the disc. The different colours show the result of the simulations with different disc aspect ratios: $h = 0.03$ (red), $h = 0.05$ (blue) and $h=0.1$ (black).}
   \label{fig:dissip_vertical_structure_time}
\end{figure*}

As mentioned in the introduction, the theory of star formation
suggests that protoplanetary discs may start their evolution from a
rather highly magnetized state, although it is not clear at present
how much magnetic flux can escape during the collapse. The relevance
of the stationary magnetic-field profile found in the previous section
depends on the time it takes for the magnetic field to evolve from
this highly magnetized initial state to its stationary state. In this
section, we therefore study the time-evolution of the magnetic-field
profile from a highly magnetized initial state. For this purpose, we
initialize the magnetic-field profile so that the parameter $\beta_0$
is uniform and equal to a rather low value, i.e.\ either $5$ or $100$
in the two cases presented below. The initial magnetic-field profile
is then obtained from equation~(\ref{eq:Bz_renormalization}). Note, however, that we do not consider discs that are so
  strongly magnetized that the MRI is suppressed, or which deviate
  strongly from Keplerian rotation.

The time-evolution of the magnetic-field profile is shown in Figure~\ref{fig:profiles_time_dissip_vertical_structure} for an initial magnetic field such that $\beta_0=5$ uniformly in the disc, an aspect ratio of $h=0.05$ and a background magnetic field which would correspond to $\beta_0=10^5$ at the outer edge of the disc. The initial magnetic field being stronger than the equilibrium value that allows a uniform value of $\beta_0$ in the stationary state, the magnetic field diffuses outwards and as a consequence decreases towards its equilibrium value (which corresponds to $\beta_0 \simeq 2\times10^5$). In a similar way as was found in Section~\ref{sec:lubow_time_evolution_diffusion}, the magnetic field diffuses outwards first in the inner parts of the disc, and, as time progresses, at larger and larger radii. 

Figure~\ref{fig:dissip_vertical_structure_time} shows the magnetic field strength at the inner edge of the disc and the magnetic flux function at the outer edge of the disc as a function of time, for three disc aspect ratios: $0.03$, $0.05$ and $0.1$. For each of these aspect ratios the background magnetic field was set so that the corresponding $\beta_0$ at the outer edge of the disc is close to its equilibrium value (i.e.\ $\beta_0=10^4$ for $h=0.1$, $\beta_0=10^5$ for $h=0.05$ and $\beta_0=10^7$ for $h=0.03$). In Section~\ref{sec:lubow_time_evolution_diffusion}, we showed that the magnetic field could diffuse outwards in a time corresponding to a diffusive timescale $t_{\rm diff}(r) \equiv \frac{r}{v_{\rm diff}} $, which is shorter than the viscous timescale by a factor of order $H/r$ (if $\Pm=1$ as assumed here). Since the transport velocities of \citet{guilet12} are in rough agreement with those of \citet{lubow94a} for strong magnetic fields, this result should hold here at least for the first stages of the evolution. In agreement with this expectation, Figure~\ref{fig:dissip_vertical_structure_time} indeed shows that the magnetic flux at the outer edge of the disc is decreased by a factor of $2.5$ in a diffusive timescale amounting to $3$ to $10\%$ of the viscous timescale, with the expected dependence on the aspect ratio. It also shows, however, that the stationary state is reached after a longer time, which is comparable to the viscous timescale. This longer timescale comes from several reasons. First, the magnetic field and magnetic flux need to decrease by two to three orders of magnitudes to reach the stationary state. This large factor therefore requires several diffusive timescales. Secondly, as the magnetic field decreases the diffusion velocity also decreases by a factor of up to $3$, leading to a longer diffusion time.

We also performed the same calculations as presented above with an initial value of $\beta_0=100$. We found very similar results, the main difference being that the time it takes to decrease the magnetic flux by a factor of a few is somewhat longer (by about $50\%$) owing to the fact that the diffusion velocity is slightly smaller for this magnetic field strength. On the other hand, the time needed to reach the stationary state is very similar to the case with an initial value of $\beta_0=5$.

Finally, we note that the results presented above may be expected to be affected by the vertical profile of the diffusion coefficients assumed in the calculation of the transport velocities. Indeed, \citet{guilet13b} showed that with a vertical dependence corresponding to either a fully turbulent disc or a disc containing a dead zone, the ratio of advection to diffusion velocities of the magnetic field was not much affected, but that the ratio of these velocities to the advection velocity of mass (relevant to the lifetime of the protoplanetary disc) was increased by a factor $\sim 2$ for a fully turbulent disc and by a factor $\sim7$ for their disc model containing a dead zone (note that the latter ratio can reach even larger values if the dead zone is more extended). With these transport velocities the stationary state may therefore be expected to be reached in a smaller fraction of the viscous timescale that sets the lifetime of the protoplanetary disc. We did not perform the same calculations as presented above with the transport velocities computed by \citet{guilet13b}, because these were computed only for weakly magnetized discs as the vertical profile of the diffusion coefficients in a highly magnetized disc is very uncertain.


\section{Discussion and conclusion}
	\label{sec:conclusion}
\subsection{Summary of the results}
We studied the global structure of the poloidal magnetic field in an accretion disc, with a particular attention to protoplanetary discs. We first used the simple transport rates of the magnetic flux usually assumed in the literature \citep{lubow94a}, which come from a crude kinematic vertical averaging and which are therefore independent of magnetic field strength. If a realistic turbulent magnetic Prandtl number of order unity is used these transport rates do not allow a significant advection of the magnetic field. Varying the magnetic Prandtl number to larger values nevertheless allows us to study in a simple way general properties of the magnetic-field structure and evolution when advection is efficient. We find that the magnetic-field profile evolves towards a stationary state that is independent of the initial magnetic-field profile in the disc (but depends on the assumed uniform ambient magnetic field, i.e.\ the strength of the interstellar magnetic field). It is also independent of the radial profile of effective viscosity (or $\alpha$ parameter) and depends only on the ratio of advection to diffusion velocities (which is here given by ${\textstyle\frac{3}{2}}\Pm h$, with $\Pm$ being the magnetic Prandtl number and $h$ the aspect ratio of the disc). In our simple disc model with uniform aspect ratio and magnetic Prandtl number, this stationary profile is well described by a self-similar solution with a power-law dependence of the magnetic field strength. The ratio of advection to diffusion velocities sets the power-law index of the magnetic-field profile varying between $b=0$ for an inefficient advection when $\Pm h \ll 1$ to $b=-2$ for a very efficient advection when $\Pm h \gg 1$. In this limit of very efficient advection our results are therefore consistent with those of \citet{okuzumi13} with a maximum magnetic field scaling like $r^{-2}$.

The normalization of this profile is set by the uniform ambient magnetic field. The magnetic field strength at the outer edge of the disc is always lower or comparable to the strength of the ambient magnetic field, while the total magnetic flux threading the disc is a factor of $1$--$2$ larger than that coming from the ambient magnetic field. The fact that the advection of magnetic field by the disc cannot increase the magnetic field at the outer edge of the disc can be understood by noting that a current ring in the disc tends to increase the magnetic field at smaller radii but to {\rm decrease} it at larger radii. At the outer edge of the disc, the currents in the disc (located at smaller radii) therefore tend to decrease the magnetic field with respect to its interstellar value. We also studied the time-evolution of the magnetic-field profile towards the steady state. Starting from an initial profile shallower than the equilibrium profile, the steady state is reached in an advective timescale (i.e.\ viscous timescale in this simple model), while an initially steeper magnetic-field profile is diffused in a resistive timescale, which is shorter than the viscous timescale by a factor $\Pm h$.

We then used the more realistic transport velocities of the magnetic flux that were computed by \citet{guilet12}. These transport velocities take into account the vertical structure of the disc as well as the back-reaction of the magnetic field on the flow. Because they are not kinematic, they depend on the strength of the magnetic field in contrast to the simple model studied before. When the magnetic pressure is comparable to the midplane thermal pressure these transport velocities agree with those of \citet{lubow94a}. For lower magnetic field strength, however, the advection velocity increases reaching up to 10 times higher than the advection velocity of mass, while the diffusion rate decreases by a factor of up to 4. This difference comes from the fact that the advection velocity of the magnetic flux is a conductivity-weighted average (rather than a density-weighted average for the advection of mass) and is therefore strongly affected by large radial velocities occurring at a height where the density is low. The dependence on the magnetic field strength is due to the fact that this average should be taken up to the height where the magnetic pressure equals the thermal pressure, which increases when the magnetic field strength is decreased.  

Using these transport rates, we showed for the first time that the magnetic field can be efficiently advected in a protoplanetary disc with a realistic turbulent magnetic Prandtl number of order unity. Owing to this advection, the magnetic field at the inner edge of the protoplanetary disc is found to be up to five orders of magnitudes larger than its interstellar value, when the ratio of outer to inner disc radii is $10^3$. This is only slightly less than the six orders of magnitudes expected for a very efficient advection as studied by \citet{okuzumi13}. Note that this amplification factor would be even larger if the disc were more extended. Because the transport rates depend on the magnetic field strength, the radial profile of magnetic field in a stationary state can be more complicated than the self-similar power law found in the calculation with simple transport rates. We nevertheless find a tendency of the magnetic-field profile to tend towards a self-similar configuration where the ratio of midplane thermal pressure to magnetic pressure $\beta_0$ is independent of radius. This is possible for an equilibrium value of $\beta_0$, which depends steeply on the aspect ratio of the disc and the turbulent magnetic Prandtl number. For a turbulent magnetic Prandtl number of order unity, the equilibrium value of $\beta_0$ is found to be in the range $\beta_0 \sim 10^4-10^7$ for aspect ratios typical of protoplanetary discs in the range $h = 0.03-0.1$. If the interstellar magnetic field strength (which roughly sets the magnetic field strength at the outer edge of the disc) corresponds to a value of $\beta$ at the disc outer edge larger than its equilibrium value, then the magnetic-field profile is steeper than the self-similar profile such that the equilibrium value of $\beta$ is reached at smaller radii. Conversely, if the interstellar magnetic field is larger than the equilibrium value at the outer edge of the disc, the magnetic profile is shallower, and $\beta$ increases with radius until it reaches its equilibrium value. This behaviour is due to the fact that the ratio of advection to diffusion velocities increases with decreasing magnetic field strength, such that weak fields lead to steeper magnetic-field profiles than stronger fields.

The relevance to protoplanetary discs of this steady-state magnetic-field configuration depends on the time it takes for the initial magnetic-field configuration to evolve towards its final steady state. Protoplanetary discs are likely to form in a rather strongly magnetized state, although some magnetic flux may already be lost to enable disc formation \citep{joos13}. We therefore studied the time-evolution of the magnetic field from a strongly magnetized initial condition (while the disc is kept fixed in a steady state). We find that a significant fraction of the excess magnetic flux is expelled from the disc in a resistive timescale which is significantly shorter than the viscous timescale driving the evolution of the protoplanetary disc.

\subsection{Consequences for protoplanetary discs}

Our results show that the strength of the large-scale poloidal magnetic field at the outer edge of a protoplanetary disc is roughly equal to (or slightly smaller than) the interstellar ambient magnetic field. Owing to the inward advection of magnetic flux counterbalancing its outward diffusion, the magnetic field increases radially inwards from this value. The slope of the profile is such that at smaller radii it reaches an equilibrium value of the ratio of midplane thermal pressure to magnetic pressure in the range $\beta_0 \sim 10^4-10^7$ (depending on the aspect ratio of the disc, with thicker discs leading to stronger magnetic fields). Such a magnetic field strength is rather weak in the sense that the magnetic pressure remains significantly smaller than the thermal pressure at the disc midplane. It can however have profound consequences on the dynamics of protoplanetary discs. At radii of 1 to a few AU, it could for example quench MRI turbulence and enable the launching of an outflow powerful enough to drive accretion at a rate compatible with observations, when the effects of ambipolar diffusion are taken into account \citep{bai13b,bai13c}. At larger radii of a few tens of AU where ambipolar diffusion is significant, it could by contrast foster the development of MRI turbulence \citep{simon13b}. It is remarkable that the range of equilibrium $\beta_0$ found in our analysis coincides with the values needed in these studies in order to explain the observed mass accretion rates.

One major motivation to study the large-scale magnetic field in protoplanetary discs is to explain the powerful collimated jets observed in T-Tauri star systems. The collimation is most likely caused by a large-scale magnetic field \citep[e.g.][]{cabrit07b}, and \citet{ferreira06} argued that a self-collimated extended disc wind launched by a magnetocentrifugal mechanism is needed to explain the observations. Our finding that the magnetic field strength increases radially inwards is very encouraging in this respect. But is it enough? Self-similar models of outflows magnetocentrifugally launched from a disc \citep[e.g.][]{ferreira97,casse00} require a strong magnetization of the disc with $\beta_0 \sim 1-10$. This is significantly lower than the values found in this paper, which might indicate at first sight that the magnetic field amplification we find is not sufficient for these models. However, note that because the outflow is very efficient at removing angular momentum from the disc, the surface density of such a jet-emitting disc can be several orders of magnitudes lower than that of a standard $\alpha$ disc with the same mass accretion rate \citep{combet08}. As a consequence the same magnetic field strength could correspond to a large value of $\beta_0$ in a standard accretion disc and to a low value of $\beta_0$ in a jet-emitting disc. 
In a jet-emitting disc where the jet is responsible for the angular momentum extraction from the disc, the vertical magnetic field strength depends on the mass accretion rate through:
\begin{equation}
B_{\rm jet} = 0.2\, \left(\frac{M}{M_\odot} \right)^{1/4}\left( \frac{\dot{M}}{10^{-7}M_\odot/{\rm year}}\right)^{1/2}\left(\frac{r}{1\,{\rm AU}}\right)^{-5/4} \frac{1}{q^{1/2}}\, {\rm G},
	\label{eq:B_jet}
\end{equation}
where $q$ is the ratio of azimuthal to vertical magnetic field strength at the surface of the disc. In an $\alpha$-disc model, where angular momentum is transported by a turbulent viscosity, the magnetic field strength is related to the mass accretion rate and the $\beta_0$ parameter by:
\begin{eqnarray}
B_{\alpha - \rm disc } &=& 0.15\, \left(\frac{M}{M_\odot} \right)^{1/4}\left( \frac{\dot{M}}{10^{-7}M_\odot/{\rm year}}\right)^{1/2}\left(\frac{r}{1\,{\rm AU}}\right)^{-5/4} \nonumber \\
&& \left(\frac{10^{-2}}{\alpha}\right)^{1/2}  \left(\frac{0.05}{h}\right)^{1/2}  \left(\frac{10^4}{\beta_0}\right)^{1/2} \, {\rm G}.
	\label{eq:B_alphadisc}
\end{eqnarray}
This opens the possibility that a standard accretion disc with a value of $\beta_0$ comparable to the low end of equilibrium $\beta_0$ found in this study could have a transition in its inner parts to a jet-emitting disc (with a value of $\beta_0$ of order unity). Combining the two above equations, the ratio of the magnetic field strength in a jet-emitting disc to that in the outer $\alpha$-disc with the same accretion rate is:
\begin{equation}
\frac{B_{\rm jet}}{B_{\alpha-{\rm disc}}} = \frac{3\alpha h \beta_0}{8 q},
	\label{eq:B_ratio_jet_alpha-disc}
\end{equation}
This ratio can equal 1 (as required at the transition) for $\beta_0=10^4$ in the outer disc, and plausible parameter values of $\alpha=10^{-2}$, $h=0.05$ and $q=2$. Note that, for $h=0.05$ as typical of the inner parts of protoplanetary discs, we have found in Section~\ref{sec:vertical_structure_stationary} an equilibrium value of $\beta_0\simeq 2.10^5$, corresponding to a magnetic field strength $4-5$ times weaker than required for this transition to a jet emitting disc. However, the right equilibrium field strength would be obtained by assuming a turbulent magnetic Prandtl number of $\Pm=2$ (instead of 1), since $\beta_0=10^4$ is obtained for $\Pm h=0.1$ (Figure~\ref{fig:betaeq_h}). Such a value of $\Pm$ is plausible, as numerical simulations suggest that the turbulent magnetic Prandtl number is of order unity within a factor of a few \citep{lesur09,fromang09,guan09}. The magnetic field strength required for a transition to a jet-emitting disc in the inner parts of the protoplanetary disc can therefore be obtained within the uncertainties of our model. This scenario should be studied further in a model that includes the outflow and its feedback on accretion. An alternative scenario is that an outflow could be launched from a rather weakly magnetised disc as suggested by local disc models \citep{suzuki09,fromang13,bai13a,bai13b}. Whether such outflows can explain the observations remains, however, to be established.

Observational constraints on the magnetic field in protoplanetary discs can come from the polarisation (or lack of it) of submillimeter emission. Above a critical magnetic field strength, dust grains are indeed expected to align with magnetic field lines and therefore to emit polarised radiation \citep{lazarian07,cho07}. Despite early claims of such detection \citep{tamura99}, recent observations have so far been unable to detect any polarisation and have put stringent constraints on the polarisation level coming from protoplanetary T-Tauri and Herbig Ae/Be discs \citep{hughes09,hughes13,krejny09}. This lack of detection could be explained either by an inefficient grain alignment, a disordered magnetic field structure or a weak magnetic field strength \citep{hughes09}. We note that the rather low magnetic field strength we predict in the outer parts of protoplanetary discs (due to the efficient diffusion of a potentially strong initial field) is well below the critical strength for grain alignment of $10-100\, {\rm mG}$ estimated by \citet{hughes09}. This may be part of the explanation for the lack of polarisation detection in T-Tauri and Herbig Ae/Be discs. The recent detection of polarisation from a young embedded disc (in the class 0 protostellar system IRAS 16293-2422) may be an indication that the magnetic field strength does indeed decrease with time \citep{rao14}. Note, however, that our analysis only predicts the mean poloidal magnetic field strength, and that a stronger potentially disordered magnetic field may well be present in the disc. Sensitive, higher resolution observations with ALMA could shed more light on this issue by resolving the disc scale height and/or probing inner regions of the disc.

The lifetime of a protoplanetary disc is set by the viscous timescale at the outer edge of the disc, or the timescale at which an outflow can drive accretion if such an outflow can be launched at these large radii. Both of these processes should depend on the strength of the large-scale poloidal magnetic field at the outer edge of the disc. Indeed, the viscous timescale depends on the strength of MRI driven turbulence, which is enhanced by the presence of such a field \citep{simon13b}, and the presence of an outflow removing angular momentum requires a large-scale magnetic field. Our results show that advection of magnetic field cannot increase the magnetic field strength at the outer edge of the disc, which remains weaker or comparable to the interstellar ambient magnetic field. We also found that an excess magnetic flux initially present in the disc would be diffused away in a timescale shorter than the lifetime of the disc. This suggests that the lifetime of a protoplanetary disc should be anticorrelated with the strength of the ambient interstellar magnetic field surrounding it (independently of the initial magnetic field in the disc), with stronger magnetic field leading to shorter lifetime. 

\citet{armitage13} have proposed a model to explain the observed two-timescale dispersal of protoplanetary discs, which relies on the evolution of a disc in the presence of a large-scale poloidal magnetic field. In this scenario, the disc first spreads viscously due the action of MHD turbulence. In a second step, a magnetically driven outflow disperses the disc in a much shorter timescale, once the surface density has decreased enough for the outflow to have a significant impact. A crucial ingredient of this scenario is the time-evolution and radial profile of the large-scale poloidal magnetic field. \citet{armitage13} did not compute it consistently, but deduced it from two simplifying assumptions: the total magnetic flux enclosed inside the disc is conserved during evolution, and the radial profile of magnetic field strength is such that the ratio of the midplane thermal pressure to magnetic pressure is uniform (though the effect of relaxing this second assumption was also studied). Our results provide some support for the second assumption, albeit mostly if the ambient interstellar magnetic field corresponds to the equilibrium value of $\beta$ at the outer edge of the disc. On the other hand, our analysis contradicts the first assumption. According to our results, an excess magnetic flux initially present in the disc should diffuse away in a timescale that is shorter than the viscous timescale. Once the magnetic field has reached a steady state (for a given disc profile at a given time), the total magnetic flux enclosed within the disc should then be between 1 and 2 times the flux of the uniform ambient interstellar field at the outer edge of the disc (which evolves with time). While this raises questions about the quantitative results of \citet{armitage13}, it does not disqualify the scenario they envisage. According to our results, once the magnetic-field configuration has reached a quasi-steady state, the magnetic field at the outer edge of the disc should be approximately independent of time, and as the surface density decreases with time due to viscous spreading the value of $\beta$ should decrease, and ultimately a magnetically driven outflow might indeed be able to drive a fast dispersal. It would be interesting to study this scenario more quantitatively by considering a time-evolving protoplanetary disc model with the magnetic-field evolution computed like in this paper.

Last but not least, we stress that our analysis is restricted by a number of simplifying assumptions, the validity of which should be quantitatively checked in the future. Firstly, our mean-field analysis relies on a very simple description of turbulence by isotropic effective diffusion coefficients (viscosity and resistivity). The effects of turbulence may be much more complex than this, with for example anisotropic diffusion as well a dynamo $\alpha$ effect that could impact our results and may enable a large-scale dynamo in addition to the advection-diffusion picture studied here. Secondly, we assumed that no outflow was launched from the disc. The presence of an outflow could change our results in several ways. It could play an important role in driving accretion \citep{bai13b,bai13c}, which may help reaching even larger magnetic fields. The presence of currents in the outflow could also change the magnetic-field structure, for example changing the inclination of the magnetic field lines at the surface of the disc for a given flux distribution. Thirdly, we considered a very simple steady-state underlying disc in order to focus on the magnetic-field evolution. A real disc should be viscously spreading with time, which can affect the magnetic-field structure \citep{okuzumi13}. Finally, we have not taken into account ambipolar diffusion and the Hall effect, which are expected to have an important impact on the outer parts of protoplanetary discs \citep{simon13b,kunz13}. Future work should take these processes into account in order to obtain a more precise description of the magnetic-field evolution.

\section*{Acknowledgments}
We thank Sylvie Cabrit and Cl\'ement Baruteau for useful discussions. JG acknowledges support from STFC grant ST/J001570/1 and from the Max-Planck--Princeton Center for Plasma Physics.


\appendix

\section{Self-similar exterior magnetic field solution}
	\label{sec:self_similar}
In this section, we derive self-similar solutions for the force-free magnetic field outside the disc, which are used in Sections~\ref{sec:lubow} and \ref{sec:vertical_structure} to interpret the numerical results. Such self-similar solutions have been found before by \citet{sakurai87} and \citet{ogilvie97}. For this derivation, we work in the spherical coordinate system $(r,\theta)$ with $\theta$ being the polar angle. In this coordinate system, equation~(\ref{eq:psi_exterior}) governing the flux function outside the disc becomes
\begin{equation}
\frac{1}{\sin\theta}\frac{\p^2 \psi}{\p r^2} + \frac{1}{r^2}\frac{\p}{\p\theta}\left(\frac{1}{\sin\theta}\frac{\p\psi}{\p\theta} \right) = 0. 
	\label{eq:force_free_field_spherical}
\end{equation}
We look for a solution of this equation of the form $\psi = f(r)g(x)$, where $x\equiv\cos\theta = z/r$. By substituting into equation~(\ref{eq:force_free_field_spherical}), one finds
\begin{equation}
r^2\frac{f^{\prime\prime}}{f} =-(1-x^2)\frac{g^{\prime\prime}}{g}. 
\end{equation}
The left-hand side depends only on $r$ while the right-hand side depends only on $x$; therefore both are equal to a constant which we call $k$.   The solutions of
\begin{equation}
r^2\frac{f^{\prime\prime}}{f} =k,
\end{equation}
are $f = Ar^{b+2}$, where $A$ is an arbitrary constant and $k=(b+1)(b+2)$. Therefore, we have found power-law solutions for all $b$ from $-\infty$ to $\infty$ when $k$ goes from $-1$ to $\infty$. However, for the flux function to not diverge when $r\rightarrow 0$, we require $b>-2$.

Then $g$ obeys the equation
\begin{equation}
(1-x^2)g^{\prime\prime} + (b+1)(b+2)g = 0.
\end{equation}
The relevant solution can be expressed in terms of the Legendre function as
\begin{equation}
g=B(1-x^2)\frac{\rmd P_{b+1}(x)}{\rmd x},
\end{equation}
where $B$ is an arbitrary constant. The 2D distribution of the flux function is then \citep{ogilvie97}:
\begin{equation}
\psi(r,x) = \psi_0 \times (1-x^2)\frac{P^\prime_{b+1}(x)}{P^\prime_{b+1}(0)}\left(\frac{r}{r_0}\right)^{b+2},
	\label{eq:flux_self_similar}
\end{equation}
where $\psi_0$ is the value of the flux function at a reference radius $r_0$ and in the disc midplane. The vertical magnetic field at the midplane reads:
\begin{equation}
B_z(r) = \frac{1}{r}\frac{\p \psi(r,0)}{\p r} = (b+2)\frac{\psi_0}{r_0^2}\left(\frac{r}{r_0}\right)^{b}.
	\label{eq:Bz_self_similar}
\end{equation}
It is therefore related to the magnetic flux at the disc midplane in the following way :
\begin{equation}
B_z(r) = (b+2)\frac{\psi(r)}{r^2}.
	\label{eq:Bz_flux_relation}
\end{equation}

The radial magnetic field is
\begin{equation}
B_{r}(r,x) =  \frac{1}{r^2\sin\theta}\frac{\p\psi}{\p \theta} =  -\frac{1}{r^2}\frac{\p\psi}{\p x}  =  \frac{\psi_0}{r_0^2}(b+1)(b+2)\frac{P_{b+1}(x)}{P^\prime_{b+1}(0)}.
	\label{eq:Br_self_similar}
\end{equation}
And finally the radial inclination of the magnetic field as $x\rightarrow 0^+$ (or indeed $z\rightarrow0^+$) is:
\begin{equation}
b_{r\mathrm{s}} \equiv \frac{B_r}{B_z}(z\rightarrow0^+) = (b+1)\frac{P_{b+1}(0)}{P_{b+1}^\prime(0)}.
	\label{eq:brs_self_similar}
\end{equation}
Note that this inclination is independent of the radius as would be expected from a self-similar solution.

\bibliography{discs}

\bsp
\label{lastpage}

\end{document}